\documentclass[preprint,12pt]{elsarticle}

\usepackage{amsmath,amssymb,mathrsfs}
\usepackage{graphicx}

\newcommand{\bra}[1]{\langle #1 |}

\newcommand{\ket}[1]{| #1\rangle}

\newcommand{\sech}{{\rm sech}}

\journal{Nuclear Physics B}

\begin{document}

\begin{frontmatter}



\title{Exotic pairing in 1D spin-3/2 atomic gases with $SO(4)$ symmetry}


\author[CSRC,Wuhan]{Yuzhu Jiang}
\author[Wuhan,Tsinghua1,Tsinghua2]{Xiwen Guan}
\ead{xwe105@wipm.ac.cn}
\author[Iopcas,X]{Junpeng Cao}
\author[CSRC]{Hai-Qing Lin}
\ead{haiqing0@csrc.ac.cn}

\address[CSRC]{Beijing Computational Science Research Center, Beijing, 100084,
China} %

\address[Wuhan]{State Key Laboratory of Magnetic Resonance and Atomic and Molecular Physics,
Wuhan Institute of Physics and Mathematics, Chinese Academy of
Sciences, Wuhan 430071, China}

\address[Tsinghua1]{Institute for Advanced Study, Tsinghua University, Beijing,
100084, China}
\address[Tsinghua2]{Department of Theoretical Physics, Research School
of Physics and Engineering, Australian National University, Canberra
ACT 0200, Australia}

\address[Iopcas]{Institute of Physics, Chinese Academy of Sciences, Beijing 100190,
China}
\address[X]{Collaborative Innovation Center of Quantum Matter, Beijing,
     China}
\begin{abstract}

Tuning interactions in the  spin singlet and quintet channels of two colliding atoms   could 
change  the symmetry of the one-dimensional  spin-3/2 fermionic systems of ultracold atoms  while preserving  the integrability.
Here we find a novel $SO(4)$ symmetry integrable point in the
spin-3/2 Fermi gas and derive  the exact solution of the model using the Bethe ansatz.
In contrast to the  model with $SU(4)$ and $SO(5)$ symmetries, the present model
with $SO(4)$ symmetry  preserves spin singlet and quintet Cooper pairs in two
sets of $SU(2)\otimes SU(2)$ spin subspaces.
We obtain full phase diagrams, including the Fulde-Ferrel-Larkin-Ovchinnikov
like pair correlations, spin excitations and quantum criticality through the
generalized Yang-Yang thermodynamic equations.
In particular, various correlation functions are calculated by using
finite-size corrections in the frame work of conformal field theory.
Moreover, within the local density approximation, we further find that spin singlet
and quintet pairs form subtle multiple shell structures in density profiles of
the trapped gas.

\end{abstract}


\end{frontmatter}

\section{Introduction}

Large-spin atomic fermions displaying  rich pairing structures and
diverse many-body phenomena could be realized through controlling
interactions in spin scattering channels.
 Promising theoretical
progress has been made on large-spin atomic fermions in the context
of multi-particle clustering superfluidity
\cite{Ho-Yi,Yi-Ho,Wu:2003,Honerkamp:2004}.
In particular, fermionic alkaline-earth atoms can exhibit an exact
$SU(\kappa)$ symmetry with $\kappa=2I+1$ \cite{Gorshkov:2010,Cazalilla}, where
$I$ is pure nuclear spin. 
Experimental explorations of these
fermionic systems have been reported on the trimer state of $^6$Li
atoms \cite{Lompe:2010,Williams:2009,Huckans:2009}, the $SU(10)$
symmetry fermionic gas of $^{87}$Sr atoms with $I=9/2$
\cite{Salvo:2010} and the two-orbital  magnetism of $SU(\kappa)$ symmetry
in $^{87}$Sr atoms \cite{Zhang:2014}, the $SU(2)\otimes SU(6)$
symmetry fermionic atoms of ${}^{173}$Yb with its spin-1/2 isotope
\cite{Taie:2010}, the $SU(6)$ Mott-insulator state of ${}^{173}$Yb
atoms \cite{Taie:2012} and two-orbital Hubbard model
\cite{Scazza:2014}, etc.
 Despite much theoretical and experimental
efforts on the large spin atomic systems, understanding  spin
pairs and large spin magnetism is still rather elusive
\cite{Krauser:2012}, see a recent review \cite{Cazallila:2014}.

In contrast to the conventional spin-1/2 electronic magnetism
\cite{Yang,Gaudin}, one-dimensional (1D) large-spin ultracold atomic
fermions exhibit richer high spin phenomena
\cite{Wu:2005,Lecheminant:2005}.
Preliminary study of multi-colour superfluidity and quartetting
ordering \cite{Capponi:2008,Roux:2009} was also carried out through few  simplest integrable large spin
fermionic systems, such as the $SU(4)$- and $SO(5)$-invariant
spin-3/2 fermions
\cite{Sutherland:1968,Guan:2009,Schlottmann:2012}.
However, there is still little progress toward to the understanding
of  large spin pairing in non-$SU(\kappa)$ symmetry models \cite{Controzzi,Jiang:2009,Rod:2010}.
To this regard, the spin-3/2 fermionic system comprises an ideal model
towards to the precise understanding of large spin non-$SU(\kappa)$
symmetries.
From experimental point of review, spin-3/2 systems can
be realized with alkali atoms of ${}^6$Li, ${}^{132}$Cs and
alkaline-earth atoms of ${}^{9}$Be, ${}^{135}$Ba and ${}^{137}$Ba
\cite{Wu:2005}, see the recent experiment \cite{Pagano} for more
details.

Advances in controlling ultracold atoms provide us promising
opportunities of realizing quantum many-body systems with more
tunable parameters \cite{Inouye, Bergeman}. For example, S-wave
scattering in these spin-3/2 atomic fermionic systems acquires six
effective interaction channels according to the Clebsch-Gordan
coefficients. One is the spin singlet channel with total spin-$0$
and the others are the spin quintet channels with total spin-$2$,
while the S-wave scattering in the channels of total spin-$1$ and
-$3$ is forbidden.
When the interactions in all scattering channels are the same, the
system exhibits $SU(4)$ symmetry and is integrable in 1D
\cite{Sutherland:1968,Guan:2009}, see model (i) in Table I.
However, if the interaction in quintet channels is different from
that in the singlet channel, then the system can exhibit a $SO(5)$
symmetry without fine tuning \cite{Wu:2005}, see model (ii) in Table
I. The model with $SO(5)$ symmetry is integrable at a special point
\cite{Jiang:2009}.
\begin{table}[ht]
\begin{center}
 \caption{Three integrable points for spin-3/2 quantum gases. \label{table1}}
 \begin{tabular}{ccccccc}
 \hline \hline
 No. & $g_{00}$ & $g_{2,2}$ & $g_{2,1}$ & $g_{2,0}$ & $g_{2,-1}$  & $g_{2,-2}$ \\
 \hline
 i & $c$ & $c$ & $c$ & $c$ & $c$ & $c$ \\
 ii & $3c$ & $c$ & $c$ & $c$ & $c$ & $c$  \\
 iii & $c$ & $-c$ & $c$ & $-c$ & $c$ & $-c$  \\
 \hline\hline
 \end{tabular}
\end{center}
\end{table}

In fact, the scattering lengths in the quintet channels may be tuned
to breakdown  the $SU(4)$ or $SO(5)$ symmetries while preserving the
integrability.
In this paper, we propose an integrable spin-3/2 atomic Fermi gas
with $SO(4)$ symmetry, see model (iii) in Table I. 
Based on the exact solution, we find that the model has subtle spin pairing
phases and exhibits Fulde-Ferrel-Larkin-Ovchinnikov (FFLO) like pair
correlations.
For $SU(4)$ symmetric model, there are
no pairs when the interactions are repulsive, while for attractive
interactions, four-body bound states appear in the ground state
\cite{Sutherland:1968,Guan:2009}.
With the competition between chemical $\mu$ and external magnetic
field $h$, three-body bound states and two-body pairs also could
emerge in the ground state \cite{Schlottmann:2012}.
%
%
However, for the present $SO(4)$ symmetry  model with either
repulsive interaction ($c>0$) or attractive interaction ($c<0$),
bound pairs could emerge in the ground state regardless of the sign
of $c$.
In contrast to the $SU(4)$ case, we also find that there are no
four- and three-body bound states in the $SO(4)$ system.
 We further show
that the integrable $SO(4)$ model presents universal quantum
criticality with dynamic exponent $z=2$ and correlation exponent
$\nu=1/2$ in terms of different pairing states.

The paper is organized as following. In Section \ref{sc:md}, we
discuss the symmetry and conserved quantities of the integrable
$SO(4)$ model in the frame work of Yang-Baxter equation.
The exact solution, thermodynamic limit and thermodynamic  Bethe
anatz equations are derived in Section \ref{sc:es}.
In Section \ref{sc:pd}, ground state properties and full phase
diagrams are discussed in grand canonical ensemble.
The elementary excitations of this model and Luttinger liquid
behaviour are studied in detail in Section \ref{sc:ee}.
In Section \ref{sc:cr}, various asymptotics of correlation functions
for several quantum phases in strong coupling limit are calculated
explicitly.
The equation of states and the quantum criticality of the system are
further studied in Section \ref{sc:qc}, followed by discussions in
Section \ref{sc:c}.

\section{ Integrable $SO(4)$ model and conserved quantities}
\label{sc:md}\setcounter{equation}{0}

We consider the model Hamiltonian
\begin{eqnarray}
 \label{H} \hat H = -\sum_{j=1}^N \frac{\partial^2}{\partial x_j^2}
 +\sum_{i\neq j}^N \sum_{lm} g_{lm} \hat P^{lm}_{ij}\delta(x_i-x_j) -
 h\hat M, \label{Ham}
\end{eqnarray}
that describes dilute spin-3/2 atomic gases of $N$ fermions with
contact interaction constrained by periodic boundary conditions to a
line of length $L$. 
Here we let  $2m=\hbar=1$. 
The last term in
the Hamiltonian (\ref{Ham}) is the Zeeman energy.
The spin polarization (magnetization) is given by $\hat M=\sum_j
\hat f^z_j$, where $\hat f^z$ is  hyperfine spin
moment in he $z$-direction. 
The projection operator $\hat P^{lm}_{ij}=\ket{lm}\bra{lm}$ projects
the total spin-$l$ state onto the spin-$m$ state in the
$z$-direction of two colliding atoms $i$ and $j$.
In the above equation,
 the summation $\sum_{lm}$ is carried out for $l=0,2$ and
$m=-l,-l+1, \cdots, l$.
The interaction strength in the channel $\ket{lm}\bra{lm}$ is given
by  $g_{lm}=-2\hbar^2/(ma_{1D}^{lm})$,
where 
$$a_{1D}^{lm}=-\frac{a_{\perp}^2}{2a^{lm} }\left(
1-C\frac{a^{lm} }{a_{\perp} }\right)$$
 is the effective 1D scattering
length depending  on the 3D scattering length $a^{lm}$  in the
channel $\ket{lm}\bra{lm}$ \cite{Olshanii1998}.
The constant  $C\approx 1.4603$ and $(l,m)$ stands for the
interaction channel.
Remarkably, the model  (\ref{Ham}) exhibits three classes of
mathematical symmetries that preserve the integrability by
appropriate choices of the interaction strength via
$c^{lm}=-2/a^{lm}_{1D}=mg_{lm}/\hbar^2$, see Table \ref{table1}.

For model (iii),  two sets of interacting channels
$\{(2,1),(0,0),(2,-1)\}$ and $\{(2,2),(2,0),(2,-2)\}$ form two spin
subspaces  with  interaction potentials $\hat V_1=\hat
P^{2,1}_{ij}+\hat P^{0,0}_{ij}+\hat P^{2,-1}_{ij}$ and $\hat
V_2=\hat P^{2,2}_{ij}+\hat P^{2,0}_{ij}+\hat P^{2,-2}_{ij}$,
respectively.
The full interaction potential is $\hat V_{ij}=c\hat V_{1}-c\hat
V_{2}$, which can also be written as {\arraycolsep=2pt
\begin{eqnarray}
 \hat V_{ij}
 =\frac12
 \left( \begin {array}{cccccccccccccccc ccc}
 0& 0& 0& 0&& 0&0& 0& 0&& 0& 0&0& 0&&0&0&0&0\\
 0&-c& 0& 0&& c&0& 0& 0&& 0& 0&0& 0&&0&0&0&0\\
 0& 0& c& 0&& 0&0& 0& 0&&-c& 0&0& 0&&0&0&0&0\\
 0& 0& 0& 0&& 0&0&-c& 0&& 0& c&0& 0&&0&0&0&0\\
 0& c& 0& 0&&-c&0& 0& 0&& 0& 0&0& 0&&0&0&0&0\\
 0& 0& 0& 0&& 0&0& 0& 0&& 0& 0&0& 0&&0&0&0&0\\
 0& 0& 0&-c&& 0&0& 0& 0&& 0& 0&0& 0&&c&0&0&0\\
 0& 0& 0& 0&& 0&0& 0& c&& 0& 0&0& 0&&0&-c&0&0\\
 0& 0&-c& 0&& 0&0& 0& 0&& c& 0&0& 0&&0&0&0&0\\
 0& 0& 0& c&& 0&0& 0& 0&& 0& 0&0& 0&&-c&0&0&0\\
 0& 0& 0& 0&& 0&0& 0& 0&& 0& 0&0& 0&&0&0&0&0\\
 0& 0& 0& 0&& 0&0& 0& 0&& 0& 0&0&-c&&0&0&c&0\\
 0& 0& 0& 0&& 0&0& c& 0&& 0&-c&0& 0&&0&0&0&0\\
 0& 0& 0& 0&& 0&0& 0&-c&& 0& 0&0& 0&&0&c&0&0\\
 0& 0& 0& 0&& 0&0& 0& 0&& 0& 0&0& c&&0&0&-c&0\\
 0& 0& 0& 0&& 0&0& 0& 0&& 0& 0&0& 0&&0&0&0&0
 \end {array} \right).
\end{eqnarray}}
Consequently, for $c>0$ (or $c<0$), the potential $\hat V_2$ (or
$\hat V_1$) presents attractive channels while $\hat V_1$ (or $\hat
V_2$) is repulsive.

The multipole operators are defined by \cite{Jiang:2009}
\begin{eqnarray}
&& \hat T^l_m =\sum_{m'=0,\pm 1} \hat T^{1}_{m'} \hat T^{l-1}_{m-m'}
C^{1,m'; l-1, m-m'}_{l,
m}, \nonumber\\[4pt]
&& m =l,l-1,\cdots, -l, \quad l = 2,3, \nonumber \\[6pt]
&& \hat T^1_{1} = -(\hat f^x + {\rm i} \hat f^y) /\sqrt{2},~ \hat
T^1_0 = \hat f^z, \quad \hat T^1_{-1} = (\hat f^x - {\rm i} \hat
f^y)/\sqrt{2},
\end{eqnarray}
where $C^{j_1,m_1;j_2,m_2}_{j,m} = \langle j_1, m_1; j_2, m_2|
j,m\rangle$ are Clebsch-Gordan coefficients and $\hat f^{x,y,z}$ are
the spin operators of  the spin-$\frac32$ atoms
\begin{eqnarray}
&& \hat f^x =
 \left(\begin{array}{cccc}
         0& \frac{\sqrt3}2&         0&         0\\
 \frac{\sqrt3}2&         0&         1&         0\\
         0&              1&         0& \frac{\sqrt3}2\\
         0&              0& \frac{\sqrt3}2&         0
 \end{array}\right),\quad
 \hat f^y =
 \left(\begin{array}{cccc}
   0& -{\rm i}\frac{\sqrt3}2&     0&     0\\
       {\rm i}\frac{\sqrt3}2&     0& -{\rm i}&    0\\
             0&      {\rm i}&     0& -{\rm i} \frac{\sqrt3}2\\
             0&            0& {\rm i}\frac{\sqrt3}2&              0
 \end{array}\right),\nonumber \\
&& \hat f^z =
 \left(\begin{array}{cccc}
 \frac32&   0&    0&    0\\
   0& \frac12&    0&    0\\
   0&   0& -\frac12&    0\\
   0&   0&    0& -\frac32
 \end{array}\right).
\end{eqnarray}
There are 15 multipole operators and only following 6 are
commutative with the interaction potential
\begin{eqnarray}
&& \hat T^{3}_{0}= \frac{3}{2\sqrt{10}}
 \left( \begin {array}{cccc}
 1&0&0&0\\  0&-3&0&0\\   0&0&3&0\\  0&0&0&-1
 \end {array} \right), \quad
 \hat T^{3}_{2}=\frac{3}{2}
 \left( \begin {array}{cccc}
  0&0&1&0\\  0&0&0&-1\\  0&0&0&0\\  0&0&0&0
 \end {array} \right), \nonumber \\
&& \hat T^{3}_{-2}=\frac32
 \left( \begin {array}{cccc}
  0&0&0&0\\ 0&0&0&0\\  1&0&0&0\\  0&-1&0&0
 \end {array} \right),
 \qquad
 \hat T^{2}_{0}=\frac12
 \left( \begin {array}{cccc}
  3&0&0&0\\  0&1&0&0\\  0&0&-1&0\\  0&0&0&-3
 \end {array} \right),\nonumber \\
&&
 \hat T^{2}_{1}=\sqrt {3}
 \left( \begin {array}{cccc}
 0&-1&0&0\\  0&0&0&0\\  0&0&0&1\\  0&0&0&0
 \end {array} \right), \qquad
 \hat T^{2}_{-1}=\sqrt{3}
 \left( \begin {array}{cccc}
  0&0&0&0\\  1&0&0&0\\  0&0&0&0\\  0&0&-1&0
 \end {array} \right).
\end{eqnarray}
Using above six multipole operators and the Pauli matrices, we define the
pseudo spin-operators as
\begin{eqnarray}
&&   j^{1,x}=\frac13(\hat T^3_{-2}+\hat T^3_2)=\frac12
\hat\sigma^x\otimes\hat\sigma^z, \quad
  j^{2,x}=\frac{\rm i}{2\sqrt{3}}
    (\hat T^2_{-1}-\hat
    T^2_1)=\frac12\hat\sigma^z\otimes\hat\sigma^x,
\nonumber     \\ &&   j^{1,y}=\frac{\rm i}3(\hat T^3_{-2}-\hat
T^3_2)
    =\frac12\hat\sigma^y\otimes\hat\sigma^z, \quad
  j^{2,y}=\frac{\rm i}{2\sqrt{3}}(\hat T^2_{-1}+\hat T^2_1)
    =\frac12\hat\sigma^z\otimes\hat\sigma^y, \nonumber \\
&&   j^{1,z}=\frac15\left(2\hat T^2_0-\frac{\sqrt{10}}3\hat
T^3_0\right)=\frac12\hat\sigma^z\otimes\hat I, \nonumber \\
&&
  j^{2,z}=\frac15\left(\hat T^2_0+\frac{2\sqrt{10}}3\hat T^3_0/3\right)
   =\frac12\hat I\otimes\hat\sigma^z.\label{pso}
\end{eqnarray}
The operators $j^{d,\alpha}$ with $d=1,2$ and $\alpha =x,\,y,\, z$
satisfy the commutation relations
\begin{eqnarray}
[j^{d,\alpha},j^{d,\beta}]={\rm i}\hbar \sum_\gamma
\epsilon_{\alpha\beta\gamma}j^{d,\gamma}, \label{pso-c}
\end{eqnarray}
where $\epsilon_{\alpha\beta\gamma}$ is the fully antisymmetric
tensor.
We clearly see that $j^{1,\alpha}$ and $j^{2,\alpha}$ generate two
$SU(2)$ Lie algebras and the six operators $j^{d,\alpha}$ form the
generators of the $SO(4)$ Lie algebra.

The attractive potentials may lead to spin-$J$ pairs with $J=0,\,2$
in the quasi-momentum space.
Consequently, the generators $\hat \phi^\dag$ of these spin-$J$
pairs comprise the following two sets of representations regarding
to the attractive channels $\hat V_2$ and $\hat V_1$
\begin{eqnarray}
&&
 \hat\phi_{2,0}=\frac{1}{\sqrt2}
 (\hat\psi_{-3/2}\hat\psi_{3/2}
 -\hat\psi_{1/2}\hat\psi_{-1/2}), \nonumber \\
&& \hat\phi_{2,2}=\hat\psi_{1/2}\hat\psi_{3/2}, \quad
 \hat\phi_{2,-2}=\hat\psi_{-3/2}\hat\psi_{-1/2},  \label{pair-J}\\[4pt]
 &&
 \hat\phi_{0,0}
 =\frac{1}{\sqrt2}(\hat\psi_{-3/2}\hat\psi_{3/2}
 +\hat\psi_{1/2}\hat\psi_{-1/2}), \nonumber \\
&& \hat\phi_{2,1}=\hat\psi_{-1/2}\hat\psi_{3/2},\quad
 \hat\phi_{2,-1}=\hat\psi_{-3/2}\hat\psi_{1/2}.  \label{pair-J-2}
\end{eqnarray}
Using the language of second quantization, the pseudo spin operators
can be expressed as $\hat j^{d,\alpha}=\hat{\boldsymbol\psi}^{\dag}
j^{d,\alpha} \hat{\boldsymbol\psi}$, where
$\hat{\boldsymbol\psi}=(\hat\psi_{3/2},\hat\psi_{1/2},\hat\psi_{-1/2},\hat\psi_{-3/2})^{\rm
t}$ and the superscript ${\rm t}$ denote  the transposition. 
One can check that
following commutation relations are valid
\begin{eqnarray}
&&[\hat j^{1\pm,z}, \hat \phi^\dag_{2,1}]=0, \quad
 [\hat j^{1\pm,z}, \hat \phi^\dag_{2,-1}]=0,\quad
 [\hat j^{1\pm,z}, \hat \phi^\dag_{0,0}]=0,\quad
\nonumber \\[4pt]
&&
 [\hat j^{2\pm,z}, \hat \phi^\dag_{2,2}]=0,\quad
 [\hat j^{2\pm,z}, \hat \phi^\dag_{2,-2}]=0,\quad
 [\hat j^{2\pm,z}, \hat \phi^\dag_{2,0}]=0.
\end{eqnarray}
The pseudo spin operators are commutative with the interaction
potential.
Therefore, the model has $SO(4)$ symmetry.
This symmetry does not preserve the number of atoms with each
internal degree of freedom. But the total number of atoms are
conserved. Meanwhile, the quantities
\begin{eqnarray}
 \hat J_{3/2}=\hat N_{3/2}-\hat N_{-3/2},~~\hat J_{1/2}=\hat N_{1/2}-\hat
 N_{-1/2},
\end{eqnarray}
are conserved. Here $\hat N_i$ is the number of atoms with spin-$i$
component.
The spin polarization along the $z$-direction, which is given by
$\hat M=\frac{3}{2}\hat J_{3/2}+\frac{1}{2}\hat J_{1/2}$, is also
conserved.

\section{The Bethe ansatz solution and thermodynamic limit}
\label{sc:es}\setcounter{equation}{0}

The model (\ref{Ham}) with $SO(4)$ symmetry can be solved by  means
of the coordinate Bethe ansatz (BA) \cite{LiebLiniger1963PR, Yang}.
The BA wave function of the model  reads
\begin{eqnarray}
 \varPsi(x_1\sigma_1, x_2\sigma_2, \cdots, x_N\sigma_N)  =\sum_{{\cal P},{\cal Q}} \varTheta({\cal Q})
  A^{\sigma_1,\sigma_2,\cdots, \sigma_N} ({\cal Q},{\cal P})
  {\rm e}^{{\rm i} \sum_{j} k_{{\cal P}_j} x_{{\cal Q}_j}}, \label{wavefunc}
\end{eqnarray}
where $x_j$ and $\sigma_j$ are the coordinate and the spin in
$z$-direction of the $j$-th atom respectively.
Here $k_j$ are pseudo-momenta of the particles,
$j=1,2,\cdots, N$,
and ${\cal Q}$ and ${\cal P}$ are the permutations of
$\{1,2,\cdots,N\}$.
$\varTheta({\cal Q})$ is a series of  production of step functions,
i.e. $\varTheta({\cal Q})=\theta(x_{{\cal Q}_2}-x_{{\cal Q}_1})
\theta(x_{{\cal Q}_3}-x_{{\cal Q}_2}) \cdots \theta(x_{{\cal
Q}_{N}}- x_{{\cal Q}_{N-1}})$.
As usual,  $\theta(x)=1$ when $x\geq1$ and $\theta(x)=0$ otherwise.
In particular, the key ingredient of BA is the superposition
coefficients of these plane waves  $\vec A({\cal Q, P})$, which can
be consequently determined by the two-body scattering relation
below.

The wave function (\ref{wavefunc}) of the many-body interacting
fermions (\ref{Ham}) should satisfy the fermionic statistics and
time independent Schr\"odinger equation $\hat H \varPsi = E\varPsi$.
These restrictions lead to  the scattering  relation between  two
superposition coefficients $\vec A({\cal Q}^{(ba)}, {\cal P}^{(ba)})
 = \hat S_{{\cal Q}_a, {\cal Q}_b} (k_{{\cal P}_a} - k_{{\cal
 P}_b})$ $
\vec A({\cal Q}^{(ab)}, {\cal P}^{(ab)})$, where the two-body
scattering matrix
\begin{eqnarray}
  S_{ab}(k)
 =\frac{k+{\rm i}c}{k-{\rm i}c} \hat V_{1,ab}
 +\frac{k-{\rm i}c}{k+{\rm i}c} \hat V_{2,ab}
 +\sum_{l=1,3} \sum_{m=-l}^l \hat P^{lm}_{ab},\label{scatt}
\end{eqnarray}
satisfies the Yang--Baxter equation
\begin{equation}
S_{ab}(\lambda) S_{ac}(\lambda+\nu) S_{bc}(\nu)= S_{bc}(\nu) S_{ac}
(\lambda+\nu) S_{ab} (\lambda),
\end{equation}
which guarantees the integrability of the model. 
In the above
equations we denoted ${\cal Q}^{(ab)} = \{\cdots,{\cal Q}_{a-1},
{\cal Q}_{a} {\cal Q}_b, {\cal Q}_{b+1}, \cdots\}$, ${\cal Q}^{(ba)}
= \{\cdots, {\cal Q}_{a-1}, {\cal Q}_b {\cal Q}_a,$ $ {\cal
Q}_{b+1}, \cdots\}$, and ${\cal P}^{(ab)}$ and ${\cal P}^{(ba)}$
have a similar definition.
Eq. (\ref{scatt}) indicates that there is no interaction in the
total spin-$1$ and -$3$ channels. This is the consequence of  the
symmetry of the wave function.
The periodic boundary conditions  give rise to  the following
eigenvalue problem
\begin{eqnarray}
 \label{eigval}
  &{\rm e}^{{\rm i} k_jL}
  \hat S_{j,j-1} (k_{j}-k_{j-1}) \cdots
 \hat S_{j,  2} (k_{j}-k_{  2})
 \hat S_{j,  1} (k_{j}-k_{  1}) \nonumber\\[4pt]
 &\hspace{35pt}\times
 \hat S_{j,  N} (k_{j}-k_{  N}) \cdots
 \hat S_{j,j+1} (k_{j}-k_{j+1}) \vec A = \vec A,
\end{eqnarray}
which is  used to determine the quasi-momenta $k$s.

\begin{figure}[t]
\begin{center}
\includegraphics[width=0.7\linewidth]{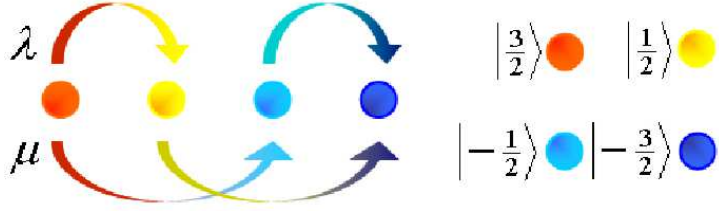}
\end{center}
\caption{ (Color online) Spin-flipping process: for  $c>0$, the spin
rapidity $\lambda$ provides the  spin flip with a total spin  change
of $1$ whereas $\nu$ gives the spin flip with a total spin change of
$2$.
For $c<0$, the rapidity $\nu$ provides the spin flip with a total
spin change of $1$ whereas $\lambda$ leads to a total spin change of
$2$. } \label{fig:lambda-nu}
\end{figure}

By using the nested algebraic BA \cite{Korepin1993book,
Martins1997NPB} and after complicated calculations, we obtain the
energy of the system as $E = \sum_{j=1}^{N} k_j^2-hM$, where the
quasi-momenta $\left\{ k_i \right\} $ should satisfy following BA
equations,
\begin{eqnarray}
&& {\rm e}^{{\rm
i}k_iL}=\prod_{j=1}^{M_1}e_{\frac12}\left(k_i-\lambda_j \right)
  \prod_{j=1}^{M_2}e_{-\frac12}\left(k_i-\nu_j \right), \,\, i=1 \cdots N, \nonumber \\
&&\prod_{i=1}^{N}e_{\frac12}\left(\lambda_j-k_i \right)
= - \prod_{\ell =1}^{M_1}e_1\left(\lambda_j-\lambda_{\ell} \right),\,\, j=1 \cdots M_1,\nonumber \\
 &&\prod_{i=1}^{N}e_{\frac12}\left(\nu_j-k_i \right)
= - \prod_{\ell =1}^{M_2}e_1\left(\nu_j-\nu_{\ell}  \right),\,\, j=1
\cdots M_2.
 \label{BAE}
\end{eqnarray}
Here $\lambda$ and $\nu$ are the spin rapidities, and we denoted the function
$e_a(x)=\frac{x-\mathrm{i} ac}{x+\mathrm{i} ac}$. 
The BA equations
coincide with the two-band model of electrons in 1D
\cite{Schlottmann92}.
From these BA equations we can obtain the  full phase diagrams and
thermodynamics of the spin-3/2 interacting fermions with $SO(4)$
symmetry. 
In Eq. (\ref{BAE}), the quasi-momenta $\left\{ k_i\right\}
$ are nested with the spin rapidities $\lambda$ and $\nu$.  For
$c>0$, the spin rapidity $\lambda$ provides the spin flip with a
total spin change of $1$ whereas $\nu$ gives the spin flip with a
total spin change of $2$, see Fig.~\ref{fig:lambda-nu}.
Thus the quantum numbers are given by $M_1=N_{1/2}+N_{-3/2}$ and
$M_2=N_{-1/2}+N_{-3/2}$ for $\left\{ \lambda \right\}$ and $\left\{
\nu\right\}$,  respectively.
For $c<0$,  the roles of the above $\nu$ and $\lambda$ are swapped.
Thus $N_{1/2}$ and $N_{-1/2}$ exchange in $M_1$ and $M_2$ in this
case.

Finding root patterns of the BA equations (\ref{BAE}) presents a big
theoretical  challenge towards to  understanding the physics of the
model.
Here we find that the quasi-momenta $\left\{k_i \right\}$ can be
either real or complex conjugated pairs at the thermodynamic limit,
$L\to \infty$ and $n=N/L$ is a constant.
The real roots $k_{{\rm u},z}\in {\mathbb{R}}$ with
$z=1,2,\cdots,N_{\rm u}$ characterize the quasi-momenta for unpaired
fermions.
The  complex conjugated roots $k_{{\rm p},z} \in \mathbb{C}$ with
$z=1,2,\cdots, N_{\rm p}$ denote the quasi-momenta of bound states,
i.e., spin-$J$ pairs with total spin  $J=0,\, 2$.
We observe that the BA equations (\ref{BAE})  present the spin pair
bound states solely depending on the attractive interaction channels
of $\hat{V}_{1}$ and $\hat{V}_2$ (or say  $c<0$ or $c>0$), see Eqs.
(\ref{pair-J}) and (\ref{pair-J-2}).
Explicitly, for $c>0$, the momenta for a bound pair of two-atom with
different spins have a complex conjugate pattern $k^{\pm}_{{\rm
p},z}=k_{{\rm p},z} \pm {\rm i}c/2$ with a binding energy
$\epsilon_{\rm p}=c^2/2$ in the thermodynamic limit.
Where a real spin rapidity  $\lambda_{{\rm p}, z}=k_{{\rm p}, z}$ is
coupled together with $k^\pm_{{\rm p}, z}$. So do the spin
rapidities $\nu_{{\rm p}, z} $ in the case of  $c<0$.

For the ground state, we observe that the BA roots only involve real
$k$, complex conjugated $k$, 1-string $\nu$ and 2-string $\nu$ for
$c>0$, see Section \ref{sc:pd}.
Here we denote the corresponding density distribution functions as
$\rho^{\rm (u)}$, $\rho^{\rm (p)}$, $\rho^{\rm (1)}$ and $\rho^{\rm
(2)}$.
Whereas, for $c<0$, the 1-string and 2-string $\lambda$  populate
the ground state instead of the $\nu$.
At finite temperatures, the spin rapidities $\lambda$'s and $\nu$'s
evolve different lengths of spin strings similar to the spin-1/2
Heisenberg integrable model \cite{Takahashi1971PTP}:
\begin{eqnarray}
  && \lambda_{n,z,j} = \lambda_{n,z} + \frac12(n+1-2j) {\rm i}|c|, \quad
  z=1,2,\cdots,M_{1,n},\nonumber\\[4pt]
 && \nu_{n,z,j} = \nu_{n,z} + \frac12(n+1-2j) {\rm i}|c|, \quad z=1,2,\cdots,M_{2,n}, \nonumber\\[4pt]
 &&
 \lambda_{n,z},\nu_{n,z} \in \mathbb{R}, \quad j=1,2,\cdots n, \label{Str}
\end{eqnarray}
where $M_{1, n}$ and $M_{2, n}$ are the numbers of $n$-string of
$\lambda$ and $\nu$,  respectively.
Thus we have  $\sum_n M_{l,n}=M_l$ ($l=1,~2$).
The BA equations in thermodynamic limit are obtained by substituting
the string hypothesis into eqs. (\ref{BAE}).
By taking the logarithm of the BA equations and using the relation
\begin{eqnarray}
\frac{x-{\rm i}c}{x+{\rm i}c}=-{\rm e}^{{\rm i} \theta_n(x)}, \quad
 \theta_n(x)=2\arctan\Big(\frac x{nc}\Big),
\end{eqnarray}
when $c>0$, we get the logarithmic form of the BA equations
\begin{eqnarray}
2\pi I^{({\rm u})}(k_{{\rm u}z})&=&
 k_{{\rm u}z}L
 -\sum_{y=1}^{N_{\rm p}}
 \theta_{\frac12}(k_{{\rm u}z}-k_{{\rm p}y})
 -\sum_{m=1}^{\infty} \sum_{y=1}^{M_{1m}}
 \theta_{\frac m2}(k_{{\rm u}z}-\lambda_{my})
 \nonumber\\
 &&  +\sum_{m=1}^{\infty} \sum_{y=1}^{M_{2m}}
 \theta_{\frac m2}(k_{{\rm u}z}-\nu_{my}),
 \nonumber\\
2\pi I^{({\rm p})}(k_{{\rm p}z})&=&
 2 k_{{\rm p}z}L
 -\sum_{y=1}^{N_{\rm p}}
 \theta_{1}(k_{{\rm p}z}-k_{{\rm p}y})
 -\sum_{y=1}^{N_{\rm u}}
 \theta_{\frac12}(k_{{\rm p}z}-k_{{\rm u}y})
 \nonumber\\
 &&
 +\sum_{m=1}^{\infty}\sum_{y=1}^{M_{2m}}
 {\mathcal A}_{1m}(k_{{\rm p}z}-\nu_{my}),
 \nonumber\\
 2\pi I^{(1)}(\lambda_{nz})&=&
 \sum_{y=1}^{N_{\rm u}}
 \theta_{\frac n2}(\lambda_{nz}-k_{{\rm u}y})
 -
 \sum_{m=1}^{\infty} \sum_{y=1}^{M_{1m}}
 {\mathcal A}_{nm}(\lambda_{nz}-\lambda_{my}),
 \nonumber\\
 2\pi I^{(2)}(\nu_{nz})
&=&\sum_{y=1}^{N_{\rm u}}
 \theta_{\frac n2}(\nu_{nz}-k_{{\rm u}y})
 + \sum_{y=1}^{N_{\rm p}}
 {\mathcal A}_{1n}(\nu_{nz}-k_{{\rm p}y})
 \nonumber\\
 &&-\sum_{m=1}^{\infty} \sum_{y=1}^{M_{1m}}
 {\mathcal A}_{nm}(\nu_{nz}-\nu_{my}), \label{lba}
\end{eqnarray}
where ${\mathcal A}_{nm}=
 \theta_{\frac{n+m}2}
 +2\big[
 \theta_{\frac{n+m-2  }2}+\cdots
 +\theta_{\frac{|n-m|-2}2}
 \big]
 +\theta_{\frac{|n-m|}2}
 -\delta_{n,m}
 \theta_{0}$.
Here the quantum number $\left\{ I^{{\rm
(u),\,(p),\,(1),\,(2)}}\right\}$ takes integers or half odd integers
as the following
\begin{eqnarray}
&&  I^{({\rm u})} \in \mathbb Z
 +\frac 12\big(N_{\rm p}+\sum_m M_{1,m} +\sum_m M_{2,m}\big), \quad I^{({\rm p})} \in \mathbb Z
 +\frac 12\big(1+N_{\rm u} +N_{\rm p}\big),\nonumber \\
&&  I^{(1)} \in \mathbb Z
 +\frac 12\big(1+n+N_{\rm u}\big),\quad I^{(2)} \in \mathbb Z
 +\frac 12\big(1+n+N\big).
\end{eqnarray}

The BA wave function acquires that any two rapidities of each branch
are not  equal. Otherwise the wave function is zero.
According to this restriction, the quantum numbers $I$s of each
branch take different numbers in parameter space
\cite{Korepin1993book}.
%
%
The solutions of the BA equations (\ref{BAE}) are classified by a
set of quantum number $\left\{ I^{{\rm
(u),\,(p),\,(1),\,(2)}}\right\}$.
Thus the momentum of the BA eigenstate  is given by
\begin{eqnarray}
 P&=&\frac{2\pi}L \left[
 \sum_{z}  I^{({\rm u})}(k_{{\rm u},z})
 +\sum_{z} I^{({\rm p})}(k_{{\rm p},z}) \right. \nonumber \\
 && \left.
 +\sum_{nz}I^{(1)}(\lambda_{n,z})
 +\sum_{nz}I^{(2)}(\nu_{n,z})
 \right].
\end{eqnarray}
Accordingly, the energy $E$ and the magnetization $M$ of the model
are given by
\begin{eqnarray}
 &&E
  = \sum_{j=1}^{N_{\rm u}} \big(k_{{\rm u},z}\big)^2
  +2\sum_{j=1}^{N_{\rm p}} \big(k_{{\rm p},z}\big)^2
  +N_{\rm p}\epsilon_{\rm p} - hM,\\
 &&
 M = \frac32N_{\rm u} +3N_{\rm p} -\sum_n (M_{1,n}+2M_{2,n}),
\end{eqnarray}
respectively. 

At finite temperature, some quantum numbers $I$s (called vacancies)
are occupied in parameter spaces whereas some of  those quantum
numbers are not occupied.
The unoccupied BA roots are called unoccupied vacancies.
In spin sector,  the unoccupied vacancies  are regarded as holes of
the spin strings.
The spin strings are nothing but spin wave  bound states.
For thermodynamic limit, i.e. $L, N \to {\infty}$ and $N/L$ is
finite, the strings and holes can be treated as density distribution
functions  $\rho(k)$ and $\rho_{\rm h}(k)$, respectively.
It follows that  $\frac{{\rm d} }{{\rm d}k}\frac{I(k)}L =\rho(k) +
\rho_{\rm h}(k)$.
Consequently,  we obtain  the integral form of the BA equations in
the thermodynamic limit,
\begin{eqnarray}
& & \rho_{\rm h}^{\rm u}+\rho^{\rm u} = \frac1{2\pi} -\hat
a_{\frac12}*\rho^{\rm p}
 - \sum_m \hat a_{\frac m2}* \big[\rho^{1,m}-\rho^{2,m}\big], \nonumber \\
 && \rho_{\rm h}^{\rm p}+\rho^{\rm p}
 = \frac1{\pi} -\hat a_{\frac12} *\rho^{\rm c}
 -\hat a_1 *\rho^{\rm p}+ \sum_m \hat A_{1,m}*\rho^{2,m},\nonumber \\
 && \rho_{\rm h}^{1n}+ \rho^{1,n} = \hat a_{\frac n2}*\rho^{\rm u}
 - \sum_m \hat A'_{nm}* \rho^{1,m},\nonumber \\
 && \rho_{\rm h}^{2n}+\rho^{2,n}
 = \hat a_{\frac n2}*\rho^{\rm u}
 + \hat A'_{1n}* \rho^{\rm p}
 - \sum_m \hat A'_{nm}* \rho^{2,m}. \label{ibae-d}
\end{eqnarray}
Here for $c>0$, $\rho^{1,n}$ and $\rho^{2,n}$ are the string
densities of $n$-strings of $\lambda$ and $n$-strings of $\nu$,
respectively.
Whereas for $c<0$,  $\rho^{1,n}$ and $\rho^{2,n}$ are the densities
of $n$-strings of  $\nu$ and $n$-string $\lambda$, respectively.
The $*$ denote the convolution $ \hat f*g(k) =
\int_{-\infty}^{\infty} f(k-k') g(k') {\rm d} k'$.
 The integral kernels are denoted as $A'_{nm}(k)=A_{nm}(k)-\delta_{nm}\delta(k)$,
$A_{nm}=a_{|n-m|/2}+2a_{(|n-m|+2)/2}+\cdots
+2a_{(n+m-2)/2}+a_{(n+m)/2}$ and $a_n(k)=\frac{1}{\pi}
\frac{n|c|}{(nc)^2+k^2}$. The $\rho^{l\,n}$ and $\rho_{\rm
h}^{l\,n}$ are the densities of particle and hole for the length
$n$-strings in spin sector.
The energy, momentum and total spin of the Hamiltonian are  given by
\begin{eqnarray}
&& e = \frac EL=\int{\rm d}k k^2 \rho^{\rm u}(k) +\int{\rm d}k \big(2k^2-\epsilon_{\rm p}\big) \rho^{\rm p}(k) -hm, \nonumber \\
&& \frac PL = \int{\rm d}k k \rho^{\rm u}(k) + \int{\rm d}k (2k) \rho^{\rm p}(k),\nonumber \\
&& m =\frac ML = \frac32 n_{\rm u} + \alpha_2 n_{\rm
p}-\alpha_1\sum_n nm_{1n}
 -\alpha_2\sum_n nm_{2n},
\end{eqnarray}
where $\alpha_1=1, \, \alpha_2=2$ for $c>0$, whereas $\alpha_1=2,
\alpha_2=1$ for $c<0$.
We also denote $n_{\rm u,p}=\int{\rm d}k \rho^{\rm u,p}(k)$  as  the
density  of atoms in the scattering state and  the density of  the
pairs, respectively.
Moreover,  $m_{l,n}=\int{\rm d}k \rho^{l,n}(k)$ with $l=1,2$  are
the densities of $n$-strings in the spin sector.

At finite temperature $T$, following Yang and Yang's grand canonical
description \cite{Yang-Yang}, the thermodynamic Bethe ansatz (TBA)
equations can be obtained from minimization of the Gibbs free energy
$\varOmega=E-TS-\mu N$ with $S$ the entropy of the system.
Here we define the dressed energies of the charge  rapidities as $
\varepsilon^{\rm \alpha}(k) = T \ln\big[\rho^{\rm \alpha}_{\rm
h}(k)/\rho^{\rm \alpha}(k)\big]$ with  $\alpha={\rm u,p}$ for
unpaired fermions and pairs, respectively,  thus the TBA equations
are given by
\begin{eqnarray}
&& \varepsilon^{\rm u}=k^2-\mu_{\rm u}
 -\hat a_{1/2} *\varepsilon^{\rm p}_-
 +\sum_n \hat a_{n/2}*(\varepsilon_{-}^{1n}+\varepsilon_{-}^{2n}),\nonumber  \\
&& \varepsilon^{\rm p}=2k^2-\mu_{\rm p}
 - \hat a_{1/2}*\varepsilon_{-}^{\rm u}
 - \hat a_1* \varepsilon^{\rm p}_-
 +\sum_n \hat A'_{1n}* \varepsilon^{2n}_-,\nonumber \\
&& \varepsilon^{1n}=n\alpha_1h -\hat a_{n/2}*\varepsilon_-^{\rm u}
 -\sum_m\hat A'_{nm}*\varepsilon^{1m}_-,\nonumber \\
&& \varepsilon^{2n}=n\alpha_2h+\hat a_{n/2}*\varepsilon^{\rm u}_-
 +\hat A'_{1n}* \varepsilon^{\rm p}_-
 -\sum_m\hat A'_{nm}* \varepsilon^{2m}_-, \label{TBAE0}
\end{eqnarray}
where $\varepsilon_-=-\ln(1+\rho/\rho_{\rm h})$. The effective
chemical potentials of unpaired fermions and pairs are defined by
$\mu_{\rm u}=\mu+(\alpha_1+\alpha_2)h/2$ and $\mu_{\rm p}=
c^2/2+2\mu+\alpha_2h$. For $c>0$, $\varepsilon^{1,n}$ and
$\varepsilon^{2,n}$ are the dressed energies for $n$-strings of
$\lambda$ and $n$-strings of $\nu$, respectively. Whereas for $c<0$,
$\varepsilon^{1,n}$ and $\varepsilon^{2,n}$ are the dressed energies
for $n$-strings of $\nu$ and $n$-strings of $\lambda$, respectively.

The TBA equations (\ref{TBAE0}) involve infinite branches of
$\varepsilon^{l,n}$ coupled together.
This imposes a big theoretical challenge  to find the exact
solutions of the TBA equations (\ref{TBAE0}).
We observe that the dressed energy of the spin string tends to be a
constant when the spin string length increases.
The larger strings contribute the smaller energy as the temperature
decreases.
In order to discuss the solutions of the TBA equations, we present
the TBA equations (\ref{TBAE0}) in the recursive form
 \begin{eqnarray}
 &&
 \varepsilon^{\rm u} = k^2-\mu -\hat G*\varepsilon^{\rm p}_-
 +\hat G* \varepsilon^{1,1}_-  +\hat G* \varepsilon^{2,1}_-,\label{TBAEEru} \\
&&
 \varepsilon^{\rm p} = 2\left(k^2-\mu-\frac 14 c^2\right)-\varepsilon^{2,1},
\label{TBAEErp} \\
 &&
 \varepsilon^{1,1}
 =-\hat G* \varepsilon^{\rm u}_- +\hat G* \varepsilon^{1,2}_+, \label{TBAEEr11}\\
 &&
 \varepsilon^{2,1}
 =\hat G*\varepsilon^{\rm u}_- +\hat G*\varepsilon^{2,2}_+,
\label{TBAEEr21} \\
 &&
 \varepsilon^{2,2} =\hat G*\varepsilon^{\rm p}_-
 + \hat G*\varepsilon^{2,1}_+ +\hat G*\varepsilon^{2,3}_+,
 \label{TBAEEr22}\\
 &&
 \varepsilon^{1,n} =\hat G* \varepsilon^{1,n-1}_+
 +\hat G* \varepsilon^{1,n+1}_+,\quad n\geq2,
 \label{TBAEEr1n}\\
 &&
 \varepsilon^{2,n} = \hat G*\varepsilon^{2,n-1}_+
 + \hat G*\varepsilon^{2,n+1}_+,\quad n\geq3,
\label{TBAEEr2n} \\
 &&
 \lim_{n\to \infty} \frac{
 \varepsilon^{1,n}}{n} =\alpha_1h,\quad
 \lim_{n\to \infty}\frac{
 \varepsilon^{2,n}}{n} =\alpha_2h.\label{TBAEErinf}
\end{eqnarray}
Here the convolution kernel is given by $G(k)=(1/{2|c|})\sech(\pi
k/c)$ and the function $\varepsilon_+$ is  defined by
$\varepsilon_+(k) = T \ln(1+\rho_{\rm h}/\rho)$.
The thermal potential per unit length is $p=p^{\rm u}+p^{\rm p}$ and
the effective pressures are given by  $p^{\rm u,p}=-\frac{r} {2\pi}
\int_{-\infty}^{\infty}dk\varepsilon_-(k)$ with $r=1$ ($r=2$) for
the unpaired fermions (the bound pairs).
The TBA equations (\ref{TBAEEru})-(\ref{TBAEErinf}) provide not only
ground state properties at zero temperatures but also full
thermodynamics at finite temperatures.
In the following Sections, we will solve the complicated TBA
equations and discuss the physics of the spin-3/2 fermionic system
with $SO(4)$ symmetry.

\section{Quantum phase diagram and pairing signature}
\label{sc:pd}\setcounter{equation}{0}

In this section, we will study  the phase diagram and the pairing
signature of the model.
A rigorous way of  finding the ground state in grand canonical
ensemble is to take the zero temperature limit in the recursive TBA
eqs. (\ref{TBAEEru})-(\ref{TBAEErinf}).
In the limit of $T\to 0$,  we have  $\varepsilon_+(k)\geq0$ and
$\varepsilon_-(k)\leq0$.
For a nonzero magnetic field $h$ and $c>0$, from Eqs.
(\ref{TBAEErinf}), we observe that the large strings of $\lambda$
and $\nu$ always have positive dressed energies.
Therefore there are no such strings in the ground state.
From Eqs. (\ref{TBAEEr1n})-(\ref{TBAEEr2n}), it is obvious that
$\varepsilon^{1,n}$ ($n\geq2$) and $\varepsilon^{2,n}$ ($n\geq3$)
are positive so that the corresponding strings do not populate in
the ground state.
Even the magnetic field tends to zero, the dressed energies of these
strings approach  to $0^+$.
Therefore  there are still no such  strings in the ground state.
The Eq. (\ref{TBAEEr11}) also gives a nonnegative dressed energy for
the real $\lambda$ (or $\nu$) rapidity for $c>0$ (or $c<0$).
Thus the dressed energy $\varepsilon^{1,1}$ is   also excluded from
the ground state.

In the $T\to 0$ limit, the TBA equations
(\ref{TBAEEr1n})-(\ref{TBAEEr2n}) reduce to a new set of dressed
energy equations that characterize the Fermi seas of the paired
states and single atoms in terms of chemical potential $\mu$ and
magnetic field $h$.
The band filling of those Fermi seas with respect to  $\mu$ and  $h$
provide an analytical way to determine the full phase diagram in the
$\mu$-$h$ plane.
For $c>0$, the dressed energy equations involve quasi-momenta
$k_{\rm u}$ and $k_{\rm p}$, and spin rapidities $\nu_1$ and
$\nu_2$.
The corresponding density distributions of them are denoted as
$\rho^{\rm u}, \rho^{\rm p}, \rho^{(1)}$ and $\rho^{(2)}$.
For $c<0$, the dressed energy equations involve $k_{\rm u}$, $k_{\rm
p}$, $\nu_1$ and $\nu_2$.
We can define the ground state densities in a similar way as the
ones  for $c>0$.
The densities of these rapidities satisfy the following integral BA
equations
\begin{eqnarray}
 \vec \rho_{\rm h}(k)+\vec \rho(k)
 = \vec \rho_0 - \hat {\boldsymbol K}* \vec \rho(k),\label{ibaerg}
\end{eqnarray}
where $\vec \rho_{\rm h}=( \rho^{\rm u}_{\rm h},\rho^{\rm p}_{\rm
h}, \rho^{(1)}_{\rm h}, \rho^{(2)}_{\rm h})^{\rm t}$, $\vec \rho
=(\rho^{\rm u}, \rho^{\rm p}, \rho^{(1)}, \rho^{(2)})^{\rm t}$,
$\vec\rho_0= ( 1/2\pi, 1/\pi, 0, 0)^{\rm t}$.
Here the superscript ${\rm t}$ means the
transposition and integral kernel $\hat{\boldsymbol K} $ is given by 
\begin{eqnarray}
 \hat K
 =\left(
 \begin{array}{cccc}
  0&\hat a_{\frac12}&-\hat a_{\frac12}&-\hat a_1\\
 \hat a_{\frac12}&\hat a_1 &-\hat a_1&-\hat a_{\frac12}-\hat a_{\frac32}\\
 -\hat a_{\frac12}&-\hat a_1  &\hat a_1  &
 \hat a_{\frac12}+\hat a_{\frac32}\\
 -\hat a_1 &-\hat a_{\frac12}-\hat a_{\frac32}
 &\hat a_{\frac12}+\hat a_{\frac32}
 &2\hat a_1+\hat a_2
 \end{array}
\right).
\end{eqnarray}
The TBA eqs. (\ref{TBAEEru})-(\ref{TBAEErinf}) in the $T\to 0$ limit
reduce to the dressed energy equations which can be written in the
following form
\begin{eqnarray}
 \vec\varepsilon(k)
 = \vec\varepsilon_0(k) - \hat {\boldsymbol K}*
 \vec\varepsilon_-(k),\label{ibaedg}
\end{eqnarray}
where $\vec \varepsilon=( \varepsilon^{\rm u},\varepsilon^{\rm p},
\varepsilon^{(1)},\varepsilon^{(2)})^{\rm t}$, $\vec \varepsilon_-=(
\varepsilon^{\rm u}_-,\varepsilon^{\rm p}_-,
\varepsilon^{(1)}_-,\varepsilon^{(2)}_-)^{\rm t}$ and
$\vec\varepsilon_0=(k^2-\mu_1, 2k^2-\mu_2, \alpha_1h,
\alpha_2h)^{\rm t}$. These  equations are  very convenient to
analyze quantum phase diagram and quantum phase transitions in terms
of chemical potential and magnetic field.

\begin{figure}[t]
\begin{center}
\includegraphics[width=0.6\linewidth]{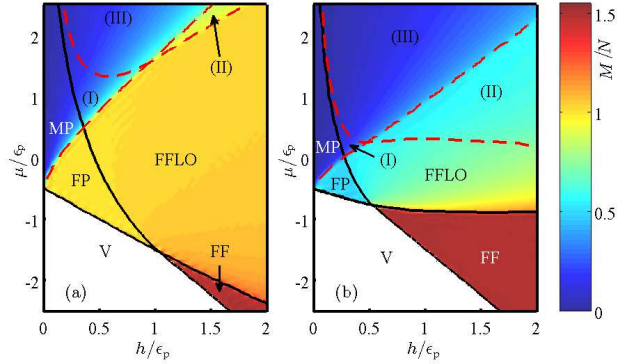}
\caption{(Color online) Phase diagram of 1D spin-3/2 Fermi gas: left
panel   for $c>0$, right panel for $c<0$.
Contour plot shows the spin polarization rate $M/N$.
The black solid lines divide the diagram into three quantum phases
in quasi-momentum space: fully-paired phase (FP), fully-polarized
phase (FF) and mixed phase of paired and unpaired fermions (FFLO).
Here V stands the vacuum state. The red dashed lines divide the
diagram into four mixed phases associated with different spin pairs, see in the text. 
Two sets of spin pair states in Eq. (\ref{pair-J}) give rise to the
pairing phases in left and right panels, respectively. }
\label{fig:phase}
\end{center}
\end{figure}

%

We observe that only the atomic pairs and the unpaired atoms
populate the ground state.
The absence of three- or four-body charge bound states is the
consequence of the repulsive interacting channels due to the $SO(4)$  symmetry.
For $c>0$, the existing pairs are $\phi^\dag_{2,\pm2}$ and
$\phi^\dag_{2,0}$, whereas for $c<0$, the paired states comprise
$\phi^\dag_{2,\pm1}$ and $\phi^\dag_{0,0}$.
The competition among the pairing, chemical potential $\mu$ and
external magnetic field $h$ gives rise to  a rich phase diagram of
this model.
Following the method presented in \cite{Guan-Ho}, full phase
diagrams of the model can be both analytically and numerically
worked out from the density equations (\ref{ibaerg}) or from the
dressed energy equations (\ref{ibaedg}). The phase diagram of the
system with $c>0$ is shown in the left panel in Fig.
\ref{fig:phase}.
For $c<0$, different BA root patterns  lead to a different phase
diagram which is presented in the  right panel in
Fig.~\ref{fig:phase}.

In the following, we only consider the regime for $c>0$.
In Fig. \ref{fig:phase}, V stands for the vacuum.
The phase FF denotes a fully-polarized phase of single $\ket{3/2}$
atoms in the region of strong magnetic field and small chemical
potential.
The FFLO-like phase composes of the largest spin-2 component pairs
$\hat\phi_{2,2}$ and excess fermions of $\ket{3/2}$ atoms that form
two mismatched Fermi surfaces in quasi-momentum space.
In this phase, atoms can be in either the scattering state
$\ket{3/2}$ or bounded pair state $\hat\phi_{2,2}$.
We show that spatial oscillation of the pair correlation function
solely depends on the mismatch between the Fermi surfaces of
$\ket{3/2}$ and $\ket{1/2}$ atoms, i.e. $\Delta k_F=\pi
(n_{3/2}-n_{1/2})$.
In the strong coupling region, the slowest decaying term of the pair
correlation function $G_{\rm
p}(x,t)=\bra{G}\hat\phi_{2,2}^{\dagger}(x,t)\hat\phi_{2,2}(0,0)\ket{G}$
is obtained, $ G_{\rm p}\approx
 {A_0\cos(\pi \Delta k_F x)}
 /({|x+{\rm i} v_{\rm u}t|^{\theta_{\rm u}}
  |x+{\rm i} v_{\rm p}t|^{\theta_{\rm p}}})$,
where $\theta_{\rm u}=1/2$, $\theta_{\rm p}=1/2+n_{\rm p}/c$ and
$n_{\rm u,p}$ are the densities of unpaired $\ket{3/2}$ atoms and
atomic pairs $\hat\phi_{2,2}$, respectively.
More detailed analysis will be further discussed in Section
\ref{sc:cr}.
Thus the momentum pair distribution has peaks at the mismatch of the
Fermi surfaces $\Delta k_F$, which presents a characteristic of the
FFLO phase.
\begin{figure}[t]
\begin{center}
\includegraphics[width=.8\linewidth]{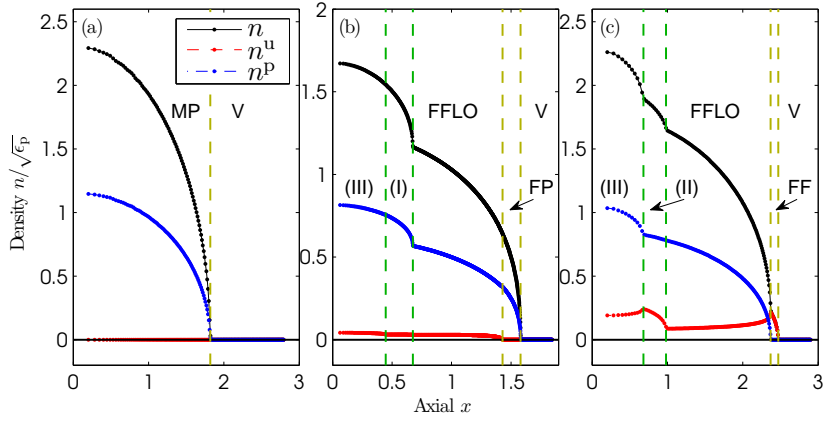}
\caption{\label{fig:LDA} (Color online) Density profiles of a
spin-imbalanced 1D Fermi gas of spin-3/2 atoms in an harmonic trap
with the setting $(h/\varepsilon_p, N/\sqrt{\varepsilon_p})=(0,
2.82), $ $(0.6, 1.75)$ and $(1.8, 3.32)$ for (a), (b) and (c),
respectively. (a) The MP  phase of $\hat\phi_{2,2},\hat\phi_{2,0}$
and $\hat\phi_{2,-2}$ pairs extends to a whole 1D tub. (b) A core of
the mixed phase III is surrounded by $6$  wing shells of mixed
states I,  FFLO state and the pure $\hat\phi_{2,2}$ paired states
FP. (c) A  core of the mixed phase III is surrounded by $6$ wing
shells of mixed states II, the FFLO state and the fully-polarised
phase FF, also see Fig.~\ref{fig:phase}. }
\end{center}
\end{figure}

Moreover, the FP phase is the fully-paired state $\hat\phi_{2,2}$,
where the pair correlation function decays as a power of distance.
The leading term of pair correlation function $G_{\rm p}$ is given
by $G_{\rm p}={A}/({|x+{\rm i} v_{\rm p}t|^{\theta_{\rm p}}})$ with
$ \theta_{\rm p}=(1/2)(1+1/\gamma)$.
This indicates that the pair correlation in the fully-paired phase
has a longer correlation length than the one  in the phase FFLO.
According to the Luttinger liquid theory, the long distance behavior
of the pair correlation of this type gives rise to the algebraic
form $G_{\rm p}\approx |x|^{-1/(2K)}$, where the parameter $K\approx
1-1/\gamma$. However, the correlation function for the one particle
Green's function decays exponentially.
%
%
When the interaction is very strong, i.e. near the phase transition
line from vacuum into the FP phase, the system behaves like a
Fermionic super Tonks-Girardeau gas \cite{ChenPRA,LMGuanPRL}.

The MP phase denotes a fully paired phase, but not a fully-polarized
$\hat\phi_{2,2}$ state.
For $c>0$, three kinds of spin pairs $\hat\phi_{2,2},\hat\phi_{2,0}$
and $\hat\phi_{2,-2}$ are involved in the MP phase.
By increasing the chemical potential $\mu$ and magnetic field $h$,
the system may enter into three other mixed phases denoted as V, FF
and FFLO Phases, see Fig. \ref{fig:phase}.
%
%
The phase transition from FFLO phase into V phase occurs as the
2-string $\nu$ emerges in the ground state.
While the one from MP phase into V phase is driven by the appearance
of real $k$, i.e. involvement of pair breaking.
The one from V phase into FFLO phase is driven by spin flipping
processes involving the real $\nu$ and the one from FF phase into
FFLO phase is driven by spin flipping process involving the 2-string
$\nu$.
%


%
In a harmonic trap, quantum  phase segments in density profiles can
be used to identify different quantum phases in experiments with
cold atoms.
We can extract the threshold values of the phase boundaries in Fig.
\ref{fig:phase} through the density profiles of the trapped gas.
Within the local density approximation, the local chemical potential
is replaced by $\mu(x)=\mu_0-\frac{1}{2}\omega_0^2x^2$ for the
trapped gas.
The total density of fermions is given by $n=n^{\rm u}+2 n^{\rm p}$.
%
%
%
For fixed particle numbers, we plot the density profiles of the
trapped gas in Fig. \ref{fig:LDA}.
In contrast to spin-1/2 Fermi gas \cite{Liao,Orso,HuiHu,Guan2007},
spin quintet pairs lead to multiple shell structures in the density
profiles.
The trapping center can be a mixture of three types of spin pairs as
well as unpaired single atoms accompanied by the different shells
involving the fully-polarized spin-2 pairs, the FFLO state, and the
FF phase, see Fig. \ref{fig:LDA}.

\section{Elementary spin and charge excitations}
\label{sc:ee} \setcounter{equation}{0}

\begin{figure}[t]
\begin{center}
 \includegraphics[width=0.9\linewidth]{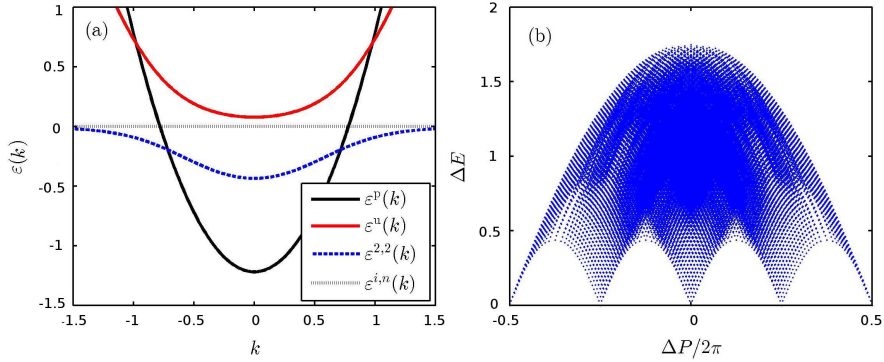}
\end{center}
\caption{\label{dreext} (Color online) 
(a) Dressed energies of different rapidities in the fully paired
phase for $h=0$ and $\gamma=1$.
The red solid line stands for  the dressed energy of the real $k$
(unpaired fermions).
The black solid line denotes  the dressed energy of the pairs
($2$-string of $k$).
The blue dashed  line shows  the dressed energy of the  2-strings of
$\nu$.
The black dotted line is the dressed energies of higher  strings of
$\nu$.
(b) Elementary excitation spectra for the model with $\gamma=1$ and
$h=0$ for the phase MP.
One $2$-string excitation of $\nu$  splits to four spin-1  holes of
$\nu$, i.e. creation of four spin-1 spinons. }
\label{fig:excitation}
\end{figure}

We consider the elementary excitations in the fully paired MP phase.
Suppose that the external magnetic field is zero.
The excitations in the charge sector of the system are similar to
the ones in the $SU(4)$ interacting fermions.
However, the spin excitations in the spin sector are quite different
from those of $SU(4)$ Fermi gas.
In the fully paired MP phase, the elementary charge excitation has
an energy gap.
In contrast to the $SU(2)$ attractive Fermi gas \cite{Guan-RMP},
there exist gapless low-lying spin excitations triggered by changing
the quantum number of 2-string $\nu$ in this paired MP phase.
The change of the number of 2-string $\nu$ results in necessary spin
configuration changes  in the paired states of
$\hat\phi_{2,2},\hat\phi_{2,0}$ and $\hat\phi_{2,-2}$.
In particularly, breaking down a 2-string $\nu$ leads to the total
spin change of $4$, i.e. creation of $4$ spinons, see Fig.~\ref{fig:excitation}.
Each spinon carries spin-1, and we denote this type of excitations
as large spinon excitations.
There is no such kind of spin excitation in the attractive $SU(2)$
Fermi gas.
These  spinons can be directly calculated from the BA equation
(\ref{ibaerg}).
One 2-string of $\nu$ is excited that leads to  four  spin-1 holes
of $\nu$: $\delta M_{2}= \delta M^{\rm h}_{2}/4$.
If one adds $d$ holes of 2-string $\nu$ into the ground state, the
excited energy and momentum are  given by
\begin{eqnarray}
\Delta E=-\sum_{j=1}^d \varepsilon^{(2)} (\nu^{\rm h}_j),\quad
\Delta P=-\sum_{j=1}^d 2\pi \int_0^{\nu^{\rm h}_j} \rho^{(2)}(\nu)
{\rm d}\nu,
\end{eqnarray}
respectively. Where $\nu^{\rm h}_j$ are the positions of the added holes.
For $h=0$ and $\gamma=c/n=1$, the 4-hole exciting spectra is shown
in Fig. \ref{dreext}(b).
However, for $c<0$,  one 2-string $\lambda$ excitation splits into
four spin-1/2 holes which give an excitation with total spin-2.

%
%

In order for our convenience to discuss correlation function,  we
will focus on three phases FF, FFLO and FP for the strong coupling
regime and a finite external magnetic field $h$.
These phases near the vacuum phase boundary in  Fig.
\ref{fig:phase}.
It is important to note that in these three phases both spin
rapidities of $\nu$ and $\lambda$ are absent in the ground state.
Thus the dressed energy equations (\ref{ibaedg}) reduce to the
similar structure as the one for the attractive spin-1/2 Fermi gas
\cite{Guan-Ho,Lee:2011a,Schlottmann:2012}.
However, spin excitations and finite temperature behaviour of these
states in the $SO(4)$ symmetry model are quite different from that
of the attractive spin-1/2 Fermi gas.
In general, at zero temperature, the Fermi points $Q_{\rm u, p}$ are
determined by the condition $\varepsilon^{\rm u}(\pm Q_{\rm u})=0$
and $\varepsilon^{\rm p}(\pm Q_{\rm p})=0$, where $Q_{\rm u, p}>0$.
At the ground state, all the quantum numbers are filled up to the Fermi
points, i.e. $I^{\rm u}(k_{{\rm u},z})-I^{\rm u}(k_{{\rm u},z-1})=1$
and $I^{\rm p}(k_{{\rm p},z})-I^{\rm p}(k_{{\rm p},z-1})=1$ with
$k_{{\rm u},z}>k_{{\rm u},z-1}$ and $k_{{\rm p},z}>k_{{\rm u},z-1}$.
All excitations can be classified as  three types of excitations,
i.e. particle-hole excitations, backwards scatting process and
adding particles at different Fermi points.
We denote these excitations as Type 1, Type II and Type III,
respectively, see Fig. \ref{fig:excitation}.

\begin{figure}[t]
\begin{center}
\includegraphics[width=0.6\linewidth]{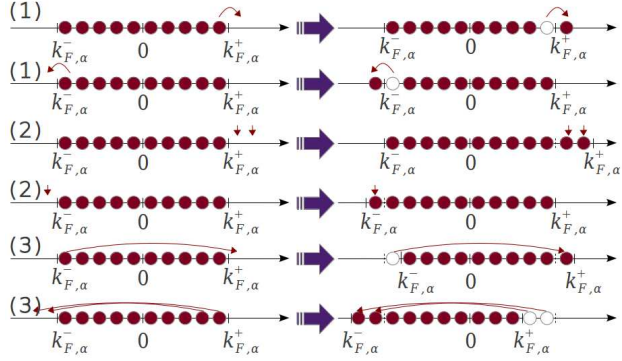}
\end{center}
\caption{\label{fig:excitation} (Color online) Three types of
elementary excitations: particle-hole excitation, adding particles
at different Fermi points and backscattering process.
$k^{\pm}_{F,\alpha}$ is the pseudo Fermi points at the right/left
Fermi points. $a={\rm u}, {\rm P}$ denote  the unpaired atoms and
atomic pairs, respectively.
The upper, middle  and lower panels show the Type I, Type II and
Type III, respectively.}
\end{figure}

The Type I elementary excitations are characterized by moving atoms
(pairs) close  the left or right pseudo Fermi point  outside  the
Fermi sea,  see Fig. \ref{fig:excitation} (1).
For example, we consider the excitations close to the right Fermi
point $Q_{\rm u}$ in real quasi-momentum space $k$.
The process is to create $N^+_{\rm u}$ holes at positions $I^{\rm
u}(k_{{\rm u},N_{\rm u}}), I^{\rm u}(k_{{\rm u},N_{\rm
u}})-1,\cdots,I^{\rm u}(k_{{\rm u},N_{\rm u}})-N^+_{\rm u}+1$ and to
add  $N^+_{\rm u}$ real $k$s outside the Fermi sea where the
corresponding quantum numbers are $I^{\rm u}(k_{{\rm u},N_{\rm
u}})+1, I^{\rm u}(k_{{\rm u},N_{\rm u}})+2,\cdots,I^{\rm u}(k_{{\rm
u},N_{\rm u}})+N^+_{\rm u}$.
Based on this setting, one can perform standard calculation of
finite-size corrections to the energy and momentum
\cite{1D-2Hubbard,Lee:2011a,Schlottmann:2012}.
This type of excitations can take place close to the left pseudo
Fermi point $-Q_{\rm u}$  with adding $N^-_{\rm u}$ holes below the
Fermi point.
In the branch of atomic pairs, particle-hole excitations also occur
close to the left and right  pseudo Fermi points with adding the
quantum numbers $N^-_{\rm p}$ and $N^+_{\rm p}$, respectively.

The Type II excitations arise from the changes of total number of
unpaired fermions or bound pairs.
This type of excitations are  characterized  by adding (or removing)
atoms/pairs  close to  the Fermi points, see Fig.
\ref{fig:excitation} (2).
We denote the change of particle number as $\Delta N_{\rm u}$ (or
$\Delta N_{\rm p}$).

The Type III excitations are created by moving particles from the
left Fermi point to the right Fermi point and vice versa. This
process  is also known as backscattering, see Fig.
\ref{fig:excitation} (3).
These two types of excitations change the pseudo Fermi points.
The quasi-momenta of single atoms/pairs with positive real parts are
viewed as the right-going atoms/pairs whereas the ones with negative
real parts are denoted as  the left-going atoms/pairs.
Both the Type II and the Type III excitations lead to the particle
number difference between the right- and left- going atoms/pairs.
We denote these number differences as $2\Delta D_{\rm u}$ and
$2\Delta D_{\rm p}$ for the unpaired atoms and the  pairs,
respectively.

Based on the above descriptions, all the three types of excitations
can be unified in the following form of the finite-size corrections
for the total  momentum and energy of elementary excitations
\begin{eqnarray}
\Delta P&=&\frac{2\pi}{L}\sum_\alpha
 \big[\Delta N_\alpha \Delta D_\alpha
 +N_\alpha \Delta D_\alpha
 +N_{\alpha}^+-N_{\alpha}^-\big], \\
\Delta E&=&\frac{2\pi}{L}\Big[
 \sum_{\alpha\beta\gamma}
 \Big(
 \frac14\Delta N_\alpha[\boldsymbol{Z}^{-1}]_{\beta\alpha}v_\beta
 [\boldsymbol{Z}^{-1}]_{\beta\gamma}\Delta N_\gamma
 +\Delta D_\alpha Z_{\alpha\beta}
 v_\beta Z_{\gamma\beta}\Delta D_\gamma\Big)
 \nonumber\\
 && +
 \sum_\alpha v_\alpha(N_{\alpha}^++N_{\alpha}^-)\Big]
 \nonumber\\
 &=&\frac{2\pi}{L}\sum_{\alpha} v_\alpha\Big(
 \frac14[\boldsymbol{Z}^{-1}\Delta \boldsymbol{N}]_\alpha^2
 +[\Delta \boldsymbol{D} \boldsymbol{Z}]_\alpha^2
 +N_{\alpha}^++N_{\alpha}^-\Big),
\end{eqnarray}
where the indices $\alpha,\beta,\gamma={\rm u},{\rm p}$ and
$v_\beta$ is the Fermi velocities of the corresponding states.
In the above equations $\boldsymbol{Z}=\boldsymbol{Z}^{(w)}$ is the
dressed charge which is determined by
\begin{eqnarray}
  \label{drec}
  {\boldsymbol{Z}}^{(w)}(k)
={\boldsymbol{I}}^{(w)}
  -\hat{\boldsymbol{K}}^{(w)}
  *{\boldsymbol{Z}}^{(w)}(k),
\end{eqnarray}
and $w$ stands for different phases, i.e. FF, FFLO and FP. For the
fully-polarised phase, $w$=FF and we have
\begin{eqnarray}
  {\boldsymbol{Z}}^{({\rm FF})}(k)
  =Z_{\rm uu}(k),\;\;
  {\boldsymbol{I}}^{({\rm FF})}=1,\;\;
  {\boldsymbol{K}}^{({\rm FF})}=0,\;\; {\boldsymbol{Z}}^{({\rm FF})}
  =Z_{\rm uu}(Q_{\rm u}).
\end{eqnarray}
For the pure paired phase, $w$=FP and we have
\begin{eqnarray}
  {\boldsymbol{Z}}^{({\rm FP})}(k)
  =Z_{\rm pp}(k),\;\;
  {\boldsymbol{I}}^{({\rm FP})}=1,\;\;
  {\boldsymbol{K}}^{({\rm FP})}(k)=a_1(k),\;\;
    {\boldsymbol{Z}}^{({\rm FP})}
  =Z_{\rm pp}(Q_{\rm p}).
\end{eqnarray}
For the phase, $w$=FFLO and we have
\begin{eqnarray}
&&
  {\boldsymbol{Z}}^{({\rm FFLO})}(k)
  =\left(\begin{array}{cc}
   Z_{\rm uu}(k)&Z_{\rm up}(k)\\
   Z_{\rm pu}(k)&Z_{\rm pp}(k)
\end{array}\right),\quad
 {\boldsymbol{K}}^{({\rm FFLO})}(k)
  =\left(\begin{array}{cc}
   0&a_{\frac12}(k)\\
   a_{\frac12}(k)&a_{1}(k)\end{array}\right),
\nonumber \\
&&  {\boldsymbol{I}}^{({\rm FFLO})}
  =\left(\begin{array}{cc} 1&0\\ 0&1 \end{array}\right),
\quad
  {\boldsymbol{Z}}^{({\rm FFLO})}
  =\left(\begin{array}{cc}
   Z_{\rm uu}(Q_{\rm u})&Z_{\rm up}(Q_{\rm u})\\
   Z_{\rm pu}(Q_{\rm p})&Z_{\rm pp}(Q_{\rm p})
\end{array}\right).
\end{eqnarray}
The value of dressed charges can be obtained by solving the dressed
charge equation (\ref{drec}) numerically.
The asymptotic forms of the dressed charges in the strong coupling
regime is given in Section \ref{sc:cr}.

On the other hand, the excited energy and excited momentum can also
be expressed by the conformal dimensions $\Delta^{\pm}$ according to
the conformal field theory \cite{Bogoliubove1990TMP, Frahm1993JPA}
\begin{eqnarray}
&& \Delta E=\frac{2\pi}{L} \sum_\alpha v_\alpha(\Delta^+_\alpha+\Delta^-_\alpha),\\
 &&
 \Delta P=\frac{2\pi}{L} \sum_\alpha (\Delta^+_\alpha-\Delta^-_\alpha)
 +2\sum_\alpha \Delta D_\alpha k_{{\rm F},\alpha},
\end{eqnarray}
where $k_{{\rm F},\alpha}$ is the Fermi momenta, $k_{{\rm
F},\alpha}= {\pi} N_\alpha/L$. It follows that
\begin{eqnarray}
&& \Delta^+_\alpha+\Delta^-_\alpha
  =N_{\alpha}^++N_{\alpha}^-
 +\frac{1}{4}[\boldsymbol{Z}^{-1}\Delta\boldsymbol{N}]_\alpha^2
 +[\Delta\boldsymbol{D Z}]_\alpha^2,
 \nonumber\\
&&
 \Delta^+_\alpha-\Delta^-_\alpha
 =N_{\alpha}^+-N_{\alpha}^-
 +\Delta N_\alpha \Delta D_\alpha.
\end{eqnarray}
These relations provide us a analytical way to calculate the
conformal dimensions via finite-size corrections
\begin{eqnarray}
 2\Delta^\pm_\alpha= 2N_{\alpha}^\pm
 \pm\Delta N_\alpha \Delta D_\alpha
 +[\Delta \boldsymbol{D} \boldsymbol{Z}]_\alpha^2
 +\frac14[\boldsymbol{Z}^{-1} \Delta \boldsymbol{N}]_\alpha^2.\label{cfmdim}
\end{eqnarray}


\section{Asymptotics of correlation functions}
\label{sc:cr}\setcounter{equation}{0}

In this section, we will calculate various correlation functions of
the system with strong interactions in three phases FF, FFLO and FP.
%
%
At zero temperature, all correlation functions reveal a power law
decay in the phases FF and FFLO.
In the pure paired FP phase the pair correlation decay as a power
law of distance whereas the single particle  Green's  function
decays exponentially.
However, at finite temperatures, all correlation functions
exponentially decay in long distance.

From the conformal field theory, the two-point correlation function
for primary fields with the conformal dimensions $\Delta^{\pm}$ is
given by \cite{Bogoliubove1990TMP}
\begin{eqnarray}
 &&G_O(x,t) =\sum
 \frac{A {\rm e}^{-2\pi{\rm i}(\sum_\alpha N_\alpha\Delta D_\alpha)x/L}}{\prod_\alpha
 (x-{\rm i}v_\alpha t)^{2\Delta^+_\alpha}
 (x+{\rm i}v_\alpha t)^{2\Delta^-_\alpha}}, \label{cr}
\end{eqnarray}
where  $G_O(x,t)=\bra{G}\hat O^{\dagger}(x,t)\hat O(0,0)\ket{G}$ is
the correlator  for  the  field  operators $\hat O(0,0)$ and  $\hat
O^{\dagger}(x,t)$.
Here the conformal dimensions $\Delta^{\pm}$  are determined by
(\ref{cfmdim}) in terms of  $N^{\pm}$, $\Delta N$, $\Delta D$ and
the dressed charge $Z$.
For strong interaction regime, the leading order of the dressed
charge is  given by $Z_{\alpha\beta}=\delta_{\alpha,\beta}$, where
$\delta_{\alpha,\beta}$ is the Kronecker delta function.
From this leading order of the dressed charge, we obtain the
conformal dimensions as $ 2\Delta_{\alpha}^\pm =
2N_{\alpha}^{\pm}+(\Delta D_{\alpha} \pm \Delta
N_{\alpha}/2)^2+O(c^{-1})$.
For the strong coupling regime, the order of $1/c$ corrections to
the correlations are presented via the conformal dimensions
\begin{eqnarray}
 2\Delta_{\alpha}^\pm = 2N_{\alpha}^{\pm}+(\Delta D_{\alpha}
\pm \Delta N_{\alpha}/2)^2+\frac{\delta_\alpha}{c}+O(c^{-2}),
\end{eqnarray}
where $\delta_\alpha$ with $\alpha=$ FF, FP and FFLO  are the first
order corrections, see Table \ref{table2}.

In the fully-polarized phase FF, fermions are unpaired and there
does not exist contact interaction.
Thus the dressed charge reads $Z^{\rm u}_{\rm u}=1$.
In the  phase FFLO, the dressed charges are given by
\begin{eqnarray}
\left(\begin{array}{cc}
  Z_{\rm uu}&Z_{\rm up}\\
  Z_{\rm pu}&Z_{\rm pp}
\end{array}\right)
=\left(\begin{array}{cc}
  1&-\frac{4Q_{\rm u}}{\pi c}\\
  -\frac{4Q_{\rm p}}{\pi  c}&1-\frac{2Q_{\rm p}}{\pi c}
\end{array}\right)+{\cal O}\Big(\frac{1}{c^3}\Big).
\end{eqnarray}
For the fully-paired  phase FP we have the dressed charge of pairs
$Z_{\rm p}^{\rm p}=1-{2Q_{\rm p}}/{\pi c}+{\cal O}(c^{-3})$.

Based on the above settings, we may calculate the asymptotics of the
correlation functions $\bra{G}\hat O^{\dagger} (x,t)\hat
O(0,0)\ket{G}$.
For a given correlation  function, the operator $O(x,t)$  act on the
ground state that  gives the quantum numbers $\Delta N_{\rm u,p}$.
Expanding the operator  $O(x,t)$ with respect to the primary fields
with conformal dimensions $\Delta^{\pm}$ and their descendent
fields, we may calculate  the asymptotics of the correlation
functions by the formula (\ref{cr}).
We present the  conformal dimensions of various correlators in
Table \ref{table2}.
In general, the real part of a correlation can be expressed as
\begin{eqnarray}
 g(x)= A\cos(2\pi x/\lambda_s)
 \prod_\alpha|x-{\rm i}v_\alpha t|^{-\theta_\alpha },
\end{eqnarray}
where $\lambda_s=(\sum_\alpha N_\alpha\Delta D_\alpha/L)^{-1}$ is
the wave length of the oscillation and the exponent $\theta_\alpha
=2\Delta^+_\alpha+2\Delta^-_\alpha$.
The Fourier transform of the correlation function near the Fermi
velocity $k_0$ is given by  \cite{Frahm1991PRB}
\begin{eqnarray}
 \tilde g(k)
\propto  [{\rm sign}(k-k_0)]^{2s}
 |k-k_0|^{\nu_s},
\end{eqnarray}
where $\nu_s=2\Delta^++2\Delta^--1$,
$s=\sum_\alpha\Delta^+_\alpha-\Delta^-_\alpha$ is the conformal spin
and $2s$ is always integer.
%
\begin{table*}\tiny
\setlength{\tabcolsep}{0pt} 
\renewcommand{\arraystretch}{1.2}
\begin{center}
 \caption{\label{table2}
 Various correlations functions for quantum phases FF, FFLO and FP.}
\begin{tabular}{ccc ccc ccc ccc ccc ccc ccc}
\hline\hline No.& CF & $N^+_{\rm u}$ & $N^-_{\rm u}$ & $\Delta
N_{\rm u}$ & $D_{\rm u}$ &  $2\Delta^+_{\rm u}$ &  $2\Delta^-_{\rm
u}$ & $\delta_{\rm u}c$ & $\theta_{\rm u}$ & $1/\lambda_s$ & $s$ &
$\nu_s$\\ \hline
1&$G^{\rm FF}_{\rm u}$ & $0$ & $0$ & $1$ & $\!-\!\frac12$ & $0$ & $1$ & $0$ & $1$ & $\!-\!\frac12 n_{\rm u}$ & $\!-\!\frac12$ & $0$\\
2&$G^{\rm FF}_{\rm u}$ & $0$ & $0$ & $1$ & $\frac12$ & $1$ & $0$ & $0$ & $1$ & $\frac12 n_{\rm u}$ & $\frac12$ & $0$\\
3&$G^{\rm FF}_{\rm u}$ & $0$ & $1$ & $1$ & $\!-\!\frac12$ & $0$ &
$3$ & $0$ & $3$ & $\!-\!\frac12 n_{\rm u}$ & $\!-\!\frac32$ & $2$\\
4&$G^{\rm FF}_{\rm n}$ & $0$ & $0$ & $0$ & $0$ & $0$ & $0$ & $0$ & $0$ & $0$ & -
 &- 
\\
5&$G^{\rm FF}_{^n}$ & $0$ & $0$ & $0$ & $1$ & $1$ & $1$ & $0$ & $2$ & $n_{\rm u}$ & $0$ & $1$\\
6&$G^{\rm FF}_{^n}$ & $0$ & $0$ & $0$ & $\!-\!1$ & $1$ & $1$ & $0$ & $2$ & $\!-\!n_{\rm n}$ & $0$ & $1$\\
7&$G^{\rm FF}_{^n}$ & $1$ & $0$ & $0$ & $0$ & $2$ & $0$ & $0$ & $2$ & $0$ & $1$ & $1$\\
8&$G^{\rm FF}_{^n}$ & $0$ & $1$ & $0$ & $0$ & $0$ & $2$ & $0$ & $2$
& $0$ & $\!-\!1$ & $1$\\ \hline\hline
No.& CF & $\begin{array}{c}_{N^+_{\rm u}}\\^{N^+_{\rm p}}\end{array}$ & $\begin{array}{c}_{N^-_{\rm u}}\\^{N^-_{\rm p}}\end{array}$ & $\begin{array}{c}_{\Delta N_{\rm u}}\\^{\Delta N_{\rm p}}\end{array}$ & $\begin{array}{c}_{D_{\rm u}}\\^{D_{\rm p}}\end{array}$ &  $2\Delta^+_{\rm u,p}$ &  $2\Delta^-_{\rm u,p}$ & $\delta_{\rm u,p}c$ & $\theta_{\rm u,p}$ & $1/\lambda_s$ & $s$ & $\nu_s$\\[-3pt] \hline
9&$G^{\rm FFLO}_{\rm u}$ & $ \begin{array}{c} 0\\
0\end{array} $ & $ \begin{array}{c} 0\\
0\end{array} $ & $ \begin{array}{c} 1\\
0\end{array} $ & $ \begin{array}{c} \frac12\\
\frac12\end{array} $ & $ \begin{array}{c} 1\!+\!\delta_{\rm u}\\
\frac14\!+\!\delta_{\rm p}\end{array} $ & $ \begin{array}{c} \delta_{\rm u}\\
\frac14\!+\!\delta_{\rm p}\end{array} $ & $ \begin{array}{c} \!-\!n_{\rm p}\\
\!-\!2 n_{\rm u}\!-\!\frac12 n_{\rm p}\end{array} $ & $ \begin{array}{c} 1\!+\!2 \delta_{\rm u}\\
\frac12\!+\!2 \delta_{\rm p}\end{array} $ & $\frac12 (n_{\rm u}\!+\!n_{\rm p})$ & $\frac12$ & $\frac12\!-\!{\frac{3 n_{\rm p}\!+\!4n_{\rm u}}{c}}$\\
10&$G^{\rm FFLO}_{\rm u}$ & $ \begin{array}{c} 0\\
0\end{array} $ & $ \begin{array}{c} 0\\
0\end{array} $ & $ \begin{array}{c} 1\\
0\end{array} $ & $ \begin{array}{c} \!-\!\frac12\\
\!-\!\frac12\end{array} $ & $ \begin{array}{c} \delta_{\rm u}\\
\frac14\!+\!\delta_{\rm p}\end{array} $ & $ \begin{array}{c} 1\!+\!\delta_{\rm u}\\
\frac14\!+\!\delta_{\rm p}\end{array} $ & $ \begin{array}{c} \!-\!n_{\rm p}\\
\!-\!2 n_{\rm u}\!-\!\frac12 n_{\rm p}\end{array} $ & $ \begin{array}{c} 1\!+\!2 \delta_{\rm u}\\
\frac12\!+\!2 \delta_{\rm p}\end{array} $ & $\!-\!\frac12(n_{\rm u}\!+\!n_{\rm p})$ & $\!-\!\frac12$ & $\frac12\!-\!{\frac {3 n_{\rm p}\!+\!4 n_{\rm u}}{c}}$\\
11&$G^{\rm FFLO}_{\rm u}$ & $ \begin{array}{c} 0\\
0\end{array} $ & $ \begin{array}{c} 0\\
0\end{array} $ & $ \begin{array}{c} 1\\
0\end{array} $ & $ \begin{array}{c} \!-\!\frac12\\
\frac12\end{array} $ & $ \begin{array}{c} \delta_{\rm u}\\
\frac14\!+\!\delta_{\rm p}\end{array} $ & $ \begin{array}{c} 1\!+\!\delta_{\rm u}\\
\frac14\!+\!\delta_{\rm p}\end{array} $ & $ \begin{array}{c} n_{\rm p}\\
2 n_{\rm u}\!-\!\frac12 n_{\rm p}\end{array} $ & $ \begin{array}{c} 1\!+\!2 \delta_{\rm u}\\
\frac12\!+\!2 \delta_{\rm p}\end{array} $ & $\frac12(n_{\rm p}\!-\!n_{\rm u})$ & $\!-\!\frac12$ & $\frac12\!+\!{\frac {n_{\rm p}\!+\!4 n_{\rm u}}{c}}$\\
12&$G^{\rm FFLO}_{\rm u}$ & $ \begin{array}{c} 0\\
0\end{array} $ & $ \begin{array}{c} 0\\
0\end{array} $ & $ \begin{array}{c} 1\\
0\end{array} $ & $ \begin{array}{c} \frac12\\
\!-\!\frac12\end{array} $ & $ \begin{array}{c} 1\!+\!\delta_{\rm u}\\
\frac14\!+\!\delta_{\rm p}\end{array} $ & $ \begin{array}{c} \delta_{\rm u}\\
\frac14\!+\!\delta_{\rm p}\end{array} $ & $ \begin{array}{c} n_{\rm p}\\
2 n_{\rm u}\!-\!\frac12 n_{\rm p}\end{array} $ & $ \begin{array}{c} 1\!+\!2 \delta_{\rm u}\\
\frac12\!+\!2 \delta_{\rm p}\end{array} $ & $\frac12 (n_{\rm
u}\!-\!n_{\rm p})$ & $\frac12$ & $\frac12\!+\!{\frac {n_{\rm
p}\!+\!4 n_{\rm u}}{c}}$\\
13&$G^{\rm FFLO}_{\rm p}$ & $ \begin{array}{c} 0\\
0\end{array} $ & $ \begin{array}{c} 0\\
0\end{array} $ & $ \begin{array}{c} 0\\
1\end{array} $ & $ \begin{array}{c} \frac12\\
0\end{array} $ & $ \begin{array}{c} \frac14\\
\frac14\!+\!\delta_{\rm p}\end{array} $ & $ \begin{array}{c} \frac14\\
\frac14\!+\!\delta_{\rm p}\end{array} $ & $ \begin{array}{c} 0\\
\frac12 n_{\rm p}\end{array} $ & $ \begin{array}{c} \frac12\\
\frac12\!+\!2 \delta_{\rm p}\end{array} $ & $\frac12 n_{\rm u}$ & $0$ & ${\frac {n_{\rm p}}{c}}$\\
14&$G^{\rm FFLO}_{\rm p}$ & $ \begin{array}{c} 0\\
0\end{array} $ & $ \begin{array}{c} 0\\
0\end{array} $ & $ \begin{array}{c} 0\\
1\end{array} $ & $ \begin{array}{c} \!-\!\frac12\\
0\end{array} $ & $ \begin{array}{c} \frac14\\
\frac14\!+\!\delta_{\rm p}\end{array} $ & $ \begin{array}{c} \frac14\\
\frac14\!+\!\delta_{\rm p}\end{array} $ & $ \begin{array}{c} 0\\
\frac12 n_{\rm p}\end{array} $ & $ \begin{array}{c} \frac12\\
\frac12\!+\!2 \delta_{\rm p}\end{array} $ & $\!-\!\frac12 n_{\rm u}$ & $0$ & ${\frac {n_{\rm p}}{c}}$\\
15&$G^{\rm FFLO}_{\rm p}$ & $ \begin{array}{c} 0\\
0\end{array} $ & $ \begin{array}{c} 0\\
0\end{array} $ & $ \begin{array}{c} 0\\
1\end{array} $ & $ \begin{array}{c} \frac12\\
1\end{array} $ & $ \begin{array}{c} \frac14\!+\!\delta_{\rm u}\\
\frac94\!+\!\delta_{\rm p}\end{array} $ & $ \begin{array}{c} \frac14\!+\!\delta_{\rm u}\\
\frac14\!+\!\delta_{\rm p}\end{array} $ & $ \begin{array}{c} \!-\!2 n_{\rm p}\\
\!-\!4 n_{\rm u}\!-\!\frac32 n_{\rm p}\end{array} $ & $ \begin{array}{c} \frac12\!+\!2 \delta_{\rm u}\\
\frac52\!+\!2 \delta_{\rm p}\end{array} $ & $\frac12 n_{\rm u}\!+\!n_{\rm p}$ & $1$ & $2\!-\!{\frac {7 n_{\rm p}\!+\!8 n_{\rm u}}{c}}$\\
16&$G^{\rm FFLO}_{\rm p}$ & $ \begin{array}{c} 1\\
0\end{array} $ & $ \begin{array}{c} 0\\
0\end{array} $ & $ \begin{array}{c} 0\\
1\end{array} $ & $ \begin{array}{c} \!-\!\frac12\\
0\end{array} $ & $ \begin{array}{c} \frac94\\
\frac14\!+\!\delta_{\rm p}\end{array} $ & $ \begin{array}{c} \frac14\\
\frac14\!+\!\delta_{\rm p}\end{array} $ & $ \begin{array}{c} 0\\
\frac12 n_{\rm p}\end{array} $ & $ \begin{array}{c} \frac52\\
\frac12\!+\!2 \delta_{\rm p}\end{array} $ & $\!-\!\frac12 n_{\rm u}$
& $1$ & $2\!+\!{\frac {n_{\rm p}}{c}}$\\
17&$G^{\rm FFLO}_{\rm n}$ & $ \begin{array}{c} 0\\
0\end{array} $ & $ \begin{array}{c} 0\\
0\end{array} $ & $ \begin{array}{c} 0\\
0\end{array} $ & $ \begin{array}{c} 0\\
0\end{array} $ & $ \begin{array}{c} 0\\
0\end{array} $ & $ \begin{array}{c} 0\\
0\end{array} $ & $ \begin{array}{c} 0\\
0\end{array} $ & $ \begin{array}{c} 0\\
0\end{array} $ & $0$ &
-
&
-
\\
18&$G^{\rm FFLO}_{n}$ & $ \begin{array}{c} 0\\
0\end{array} $ & $ \begin{array}{c} 0\\
0\end{array} $ & $ \begin{array}{c} 0\\
0\end{array} $ & $ \begin{array}{c} 0\\
1\end{array} $ & $ \begin{array}{c} 0\\
1\!+\!\delta_{\rm p}\end{array} $ & $ \begin{array}{c} 0\\
1\!+\!\delta_{\rm p}\end{array} $ & $ \begin{array}{c} 0\\
\!-\!2 n_{\rm p}\end{array} $ & $ \begin{array}{c} 0\\
2\!+\!2 \delta_{\rm p}\end{array} $ & $n_{\rm p}$ & $0$ & $1\!-\!{\frac{4n_{\rm p}}{c}}$\\
19&$G^{\rm FFLO}_{\rm n}$ & $ \begin{array}{c} 0\\
0\end{array} $ & $ \begin{array}{c} 0\\
0\end{array} $ & $ \begin{array}{c} 0\\
0\end{array} $ & $ \begin{array}{c} 0\\
\!-\!1\end{array} $ & $ \begin{array}{c} 0\\
1\!+\!\delta_{\rm p}\end{array} $ & $ \begin{array}{c} 0\\
1\!+\!\delta_{\rm p}\end{array} $ & $ \begin{array}{c} 0\\
\!-\!2 n_{\rm p}\end{array} $ & $ \begin{array}{c} 0\\
2\!+\!2 \delta_{\rm p}\end{array} $ & $\!-\!n_{\rm p}$ & $0$ & $1\!-\!{\frac{4n_{\rm p}}{c}}$\\
\hline\hline No.& CF & $N^+_{\rm p}$ & $N^-_{\rm p}$ & $\Delta
N_{\rm p}$ & $D_{\rm p}$ &  $2\Delta^+_{\rm p}$ &  $2\Delta^-_{\rm
p}$ & $\delta_{\rm p}c$ & $\theta_{\rm p}$ & $1/\lambda_s$ & $s$ &
$\nu_s$\\ \hline
24&$G^{\rm FP}_{\rm p}$ & $0$ & $0$ & $1$ & $0$ & $\frac14\!+\!\delta_{\rm p}$ & $\frac14\!+\!\delta_{\rm p}$ & $\frac12 n_{\rm p}$ & $\frac12\!+\!2 \delta_{\rm p}$ & $0$ & $0$ & $\!-\!\frac12\!+\!{\frac {n_{\rm p}}{c}}$\\
25&$G^{\rm FP}_{\rm p}$ & $0$ & $0$ & $1$ & $1$ & $\frac94\!+\!\delta_{\rm p}$ & $\frac14\!+\!\delta_{\rm p}$ & $\!-\!\frac32 n_{\rm p}$ & $\frac52\!+\!2 \delta_{\rm p}$ & $n_{\rm p}$ & $1$ & $\frac32\!-\!{\frac{3n_{\rm p}}{c}}$\\
26&$G^{\rm FP}_{\rm p}$ & $0$ & $0$ & $1$ & $\!-\!1$ & $\frac14\!+\!\delta_{\rm p}$ & $\frac94\!+\!\delta_{\rm p}$ & $\!-\!\frac32 n_{\rm p}$ & $\frac52\!+\!2 \delta_{\rm p}$ & $\!-\!n_{\rm p}$ & $\!-\!1$ & $\frac32\!-\!{\frac{3n_{\rm p}}{c}}$\\
27&$G^{\rm FP}_{\rm p}$ & $1$ & $0$ & $1$ & $0$ &
$\frac94\!+\!\delta_{\rm p}$ & $\frac14\!+\!\delta_{\rm p}$ &
$\frac12 n_{\rm p}$ & $\frac52\!+\!2 \delta_{\rm p}$ & $0$ & $1$ &
$\frac32\!+\!{\frac {n_{\rm p}}{c}}$\\
28& $G^{\rm FP}_n$ & $0$ & $0$ & $0$ & $0$ & $0$ & $0$ & $0$ & $0$ & $0$ & -
 &
-
\\
29&$G^{\rm FP}_n$ & $0$ & $0$ & $0$ & $1$ & $1\!+\!\delta_{\rm p}$ & $1\!+\!\delta_{\rm p}$ & $\!-\!2 n_{\rm p}$ & $2\!+\!2 \delta_{\rm p}$ & $n_{\rm p}$ & $0$ & $1\!-\!{\frac{4n_{\rm p}}{c}}$\\
30&$G^{\rm FP}_n$ & $0$ & $0$ & $0$ & $\!-\!1$ & $1\!+\!\delta_{\rm p}$ & $1\!+\!\delta_{\rm p}$ & $\!-\!2 n_{\rm p}$ & $2\!+\!2 \delta_{\rm p}$ & $\!-\!n_{\rm p}$ & $0$ & $1\!-\!{\frac{4n_{\rm p}}{c}}$\\
31&$G^{\rm FP}_n$ & $1$ & $0$ & $0$ & $0$ & $2$ & $0$ & $0$ & $2$ & $0$ & $1$ & $1$\\
\hline\hline
\end{tabular}
\end{center}
\end{table*}

The various correlation functions, for example, single particle
Green's function, pair-pair correlation, density-density correlation
are listed in Table \ref{table2}. The fully polarized case FF is
regarded as a single component free fermionic gas.
The correlation function  of the field operator is given by
\begin{eqnarray}
 G_{\rm u}(x,t)=\bra{G}{\hat\psi^\dag_{3/2}(x,t)\hat\psi_{3/2}(0,0)}\ket{G}. \label{cfu0}
\end{eqnarray}
This excitation associates with the quantum number $\Delta N_{\rm
u}=1$ and $\Delta D_{\rm u}\in \mathbb Z+1/2$.
A few leading orders of the asymptotics of the single particle
Green's function are obtained by choices of the values of the
$\Delta D$, see the rows 1-3 in the  Table \ref{table2}.
The leading term in the correlation function  $G_{\rm u}$ is given
by
\begin{eqnarray}
 G_{\rm u}^{\rm FF}
 \approx A_0\frac{\cos(\pi n_{\rm  u}x)}{|x-{\rm i} v_{\rm u}t|},
\end{eqnarray}
with the setting $\Delta D_{\rm u}=\pm1/2$ and $N^+_{\rm u}=N^-_{\rm
u}=0$, see the  row-1 and the row-2 in Table \ref{table2}.
 Here  $n_{\rm u}$ is the density of single  atoms and $A_0$ is some constant.
The row-1  shows  a left-going wave with $\Delta D_{\rm u}=-1/2$, or
say  that adding an unpaired atom near the left Fermi point.
The row-2 indicates a right-going excitation wave with  $\Delta
D_{\rm u}=1/2$.
We also consider the charge density correlation function $
G_{n}(x,t) = \langle n(x,t)n(0,0)\rangle $ in this phase, namely
\begin{eqnarray}
 G_{n}=
 \bra{G}{ \hat n(x,t)\hat n(0,0)}\ket{G} =n^2+\frac{A_{n1}\cos(2\pi n_{\rm u} x) }{|x-\mathrm{i} v_{\rm u}t|^2}+ \frac{A_{n2} }{|x-\mathrm{i} v_{\rm u}t|^2},
\end{eqnarray}
where $A_{n1}$ and $A_{n2}$ are constants, see the rows 4-8 in the
Table \ref{table2}.

In phase FFLO, both the spin-2 pairs and single  spin-3/2 atoms
coexist.
The single particle Green's function $G_{\rm u}$ acquires $\Delta
N_{\rm u}=1$, $\Delta N_{\rm p}=0$ and $\Delta D_{\rm u,p}\in
\mathbb{Z}+1/2$.
The few leading orders of the asymptotics of $G_{\rm u}$ are
indicated in the rows 9-12 in Table \ref{table2}. The first two
leading terms read
\begin{eqnarray}
 G^{\rm FFLO}_{\rm u} \approx
 \frac{A_{1}
 \cos[\pi(n_{\rm u}+n_{\rm p})x]}
  {|x+{\rm i} v_{\rm u}t|^{\theta_{\rm u1}}
   |x+{\rm i} v_{\rm p}t|^{\theta_{\rm p1}}}
 +\frac{A_{2}\cos[\pi(n_{\rm u}-n_{\rm p})x]}
  {|x+{\rm i} v_{\rm u}t|^{\theta_{\rm u2}}
   |x+{\rm i} v_{\rm p}t|^{\theta_{\rm p2}}}, \label{49}
\end{eqnarray}
where $A_1$ and $A_2$ are constants and the exponents read
\begin{eqnarray}
 &&\theta_{\rm u1}=1-2n_{\rm p}/c,\quad
 \theta_{\rm p1}=1/2-4n_{\rm u}/c-n_{\rm p}/c,
 \nonumber\\[4pt]
 &&\theta_{\rm u2}=1+2n_{\rm p}/c,\quad
 \theta_{\rm p2}=1/2+4n_{\rm u}/c-n_{\rm p}/c.
\end{eqnarray}
In the correlation function (\ref{49}), the first term presents the
result listed in the  rows 9 and 10 in Table \ref{table2}, and the
second term is presented in  the  rows 11 and 12.
For finitely strong interaction, the single particle correlation
function decays slower than that of the free Fermions due to the
correlations between pairs and between pairs and  unpaired atoms.

In the FFLO phase, the asymptotic  pair-pair correlation is given by
\begin{eqnarray}
 \label{cfp0}
 G_{\rm p}(x,t)=\bra{G}{\phi^\dag_{\rm p} (x,t) \phi_{\rm p}
 (0,0)}\ket{G},
\end{eqnarray}
with the quantum numbers  $\Delta N_{\rm u}=0$, $\Delta N_{\rm
p}=1$, $\Delta D_{\rm u}\in \mathbb{Z}+1/2$ and $\Delta D_{\rm p}\in
\mathbb{Z}$.
The leading order of the pair-pair correlation function is given
explicitly by
\begin{eqnarray}
G_{\rm p}\approx
 \frac{A_3\cos[\pi (n_{3/2}-n_{1/2})x]}
 {|x+{\rm i} v_{\rm u}t|^{\theta_{\rm u3}}
  |x+{\rm i} v_{\rm p}t|^{\theta_{\rm p3}}},
\end{eqnarray}
where $A_3$ is a constant and
\begin{eqnarray}
 \theta_{\rm u3}=1/2, \quad
 \theta_{\rm p3}=1/2+n_{\rm p}/c,
\end{eqnarray}
see the rows 13 and 14 in Table \ref{table2}.
Here $n_{3/2}$ and $ n_{1/2}$ are the densities of spin-3/2 and -1/2
atoms, respectively.
We see that the long distance asymptotics for the pair correlation
function oscillates with the wave number $\pi (n_{3/2}-n_{1/2})$, a
mismatch between the two Fermi surfaces   of the  two species.
This is a characteristic of the FFLO-like correlation.
We would also like to address  that the FFLO-like oscillation terms
arise from Type III excitations, i.e. backscattering for bound pairs
and unpaired fermions.
The charge density correlation function $ G_{n}(x,t)$ is presented
in the rows 17-19 in Table \ref{table2}.

In the phase FP, the  pair-pair correlation function $G_{\rm p}$
acquires the quantum number $\Delta N_{\rm p}=1$.
We present the result of the long distance asymptotics of pair-pair
correlations in the rows 24-27 in Table \ref{table2}.
The leading term of the pair-pair correction reads
\begin{eqnarray}
 \label{g4p}
 G_{\rm p} \approx\frac{A_4}{|x+{\rm i} v_{\rm p}t|^{\theta_{\rm p4}}},~~~
 \theta_{\rm p4}=\frac12 +\frac{n_{\rm p}}c,
\end{eqnarray}
where $A_4$ is a constant.
This shows that  the exponent  $\theta_{\rm p4}$ is smaller than
that in the phase FFLO.
In this phase the pair-pair correlation dominates the quantum
correlation.
Finally, we present the charge density-density correlations
\begin{eqnarray}
 G_{n}=
 \bra{G}{ \hat n(x,t)\hat n(0,0)}\ket{G} =n^2+A_{{n}2}g(x,t),
 \end{eqnarray}
with the correlation
\begin{eqnarray}
g(x,t)=\frac{\cos(2 \pi  n_{\rm p}x)}
 {|x+{\rm i} v_{\rm p}t|^{\theta_1}}, \quad \theta_1=2-4n_p/c,
\end{eqnarray}
where $A_{n2}$ is constant. The related quantum numbers are listed
in the rows 28-31 in Table \ref{table2}.

\section{Equation of state and quantum criticality}
\label{sc:qc}\setcounter{equation}{0}

The 1D fermionic systems usually exhibit a rich resource of
Luttinger liquids and show  a novel magnetism  in term of
spin-change separation scenario, see recent review \cite{Guan-RMP}.
In order to understand large spin cold atoms in 1D, it is very
important to investigate  the low temperature behavior of Luttinger
liquids.
We will prove that the Luttinger liquids of different pairing states
comprise a universal low energy physics of the model.
In general, the specific heat can be written in terms of sound
velocities of the liquid phases, i.e. $c_L=(\pi TL/3) \sum_\alpha
1/v_\alpha$, where the summation  carries out  over all the Fermi
velocities in the charge and spin sectors.
This result  is valid  for the temperature below a crossover
temperature $T^{*}$, i.e. $T \ll T^{*} $.
Here the crossover temperature can be determined from the equation
of states \cite{Guan:2013}.
Near a critical point and for the  temperature  $T\gg T^{*}$, the
system lies in  the critical regime, where the crossover temperature
characterizes the energy gap $T^{*} \sim \mu-\mu_{\rm c}$ and/or
$T^{*} \sim H-H_{\rm c}$.
In the critical regime, universal scaling behaviour of
thermodynamical properties  is expected to distinguish quantum
criticality with the critical dynamic exponent $z=2$ from the
Luttinger liquid criticality of $z=1$ \cite{Guan-Ho}.

For strong coupling regimes,  only  charge rapidities are left at
the ground states for the phases  FF, FFLO and FP.
For low temperatures, the spin wave contributions to the dressed
energies  of these phases  can be analytically derived.
In the same fashion as the $SU(2)$ Fermi gas \cite{Guan-Ho}, we can
express the TBA equations of the dressed energies  as
\begin{eqnarray}
\textstyle \varepsilon^{\alpha }(k)=r_{\alpha }k^2-\mu _{\alpha
}-\sum _{\beta}K_{\alpha \beta }*\epsilon^{\beta}_-(k)+f_{\alpha
}(k),\quad \alpha={\rm u}, {\rm p}.
\end{eqnarray}
Here $r_\alpha$ is viewed as  the effective mass of the
corresponding charges.
For the unpaired atoms, $r_{\rm u}=1$ and for the pairs, $r_{\rm
p}=2$.
In the above equations, $f_{\alpha}$ denotes  the contributions of
spin wave bound states to the dressed energies of the charge degree
of freedoms.
However, for $h\gg T$,  the spin contribution term   $f_{\alpha}$
will exponentially decay as ${\rm e}^{-h/T}$ and it is negligible in
low energy physics.
Follow the method proposed in \cite{Guan-Ho}, for  the strong
interaction regime, we find the following equation of states
\begin{eqnarray}
 p_{\alpha }=\frac{r_{\alpha }^{\frac12}T^{\frac32}}{2\pi^{\frac12}}F_{\frac12}\Big(\frac{A_{\alpha }}{T}\Big)
 \Big(1-\sum _{\beta }
 \frac{D_{\alpha \beta }^{(3)}p_{\beta }}{c^3r_{\alpha }r_{\beta
 }}\Big)+O(c^{-5}).
\end{eqnarray}
Here the effective chemical potentials $A_\alpha$  is given by 
\begin{eqnarray}
A_{\alpha }=\mu _\alpha-\sum_\beta \frac{2D_{\alpha \beta}^{(1)}}{
 r_\beta c} p_\beta -
 \sum_\beta
 \frac{D_{\alpha \beta }^{(3)}T^{\frac52}}{2r_\beta^{\frac32}\pi ^{\frac12} c^3}
 F_{\frac32}\big(\frac{A_{\beta }}{T}\big),
\end{eqnarray}
and $D^{(a)}$ are constants resulted from the strong coupling
expansion of the integral kernels $K$.
$F_{j}(A_{\alpha}/T)$ are Fermi--Dirac integrals which can also be
written as  polylogarithm functions \cite{Guan-Ho}.
$D^{(1)}=D^{(3)}=0$ in the  phase FF. In the FLLO phase, we have
\begin{eqnarray}
D^{(1)}=\left(
\begin{array}{cc}
 0 \;\;\; & 2 \\ 2\;\;\;  & 1
\end{array}
\right), \quad D^{(3)}=-\left(
\begin{array}{cc}
 0 \;\;\; & 8 \\ 8\;\;\;  & 1
\end{array}\right).
\end{eqnarray}
In the FP phase, these constants are given by  $D^{(1)}=-D^{(3)}=1$.
By iteration with $p_{\alpha}$ and the effective chemical potentials
$A_{\rm u,p}$, we obtain a close form of  the equation of sates for
the FFLO phase  as
\begin{eqnarray}
 p &=& \frac{T^{\frac32}}{\sqrt{2\pi}}
 F_{\frac12}^{\rm p}\Big\{1+\frac{4T^{\frac32}}{\sqrt{2\pi}c^3}
 \Big[F_{\frac12}^{\rm u}+\frac{1}{16}F_{\frac12}^{\rm p}\Big]\Big\}
\nonumber \\
&& +\frac{T^{\frac32}}{2\sqrt{\pi}}F_{\frac12}^{\rm u}
 \Big[1+\frac{4T^{\frac32}}{\sqrt{2\pi }c^3}
 F_{\frac12}^{\rm p}\Big]+O(c^{-5}), \label{es}
\end{eqnarray}
where $F_{j}^{\alpha}=F_{j}(A_{\alpha}/T)$ and renormalized chemical
potentials $A_{\rm u,p}$ are given by
\begin{eqnarray}
&& A_{\rm u}
 =\mu _1-\frac{2T^{3/2}}{\sqrt{2\pi}c}F_{1/2}^{\rm p}
 +\frac{2T^{5/2}}{\sqrt{2\pi}c^3}F_{3/2}^{\rm p},\nonumber \\
 &&A_{\rm p}=\mu _2
 -\frac{2T^{3/2}}{\sqrt{ \pi}c}F_{1/2}^{\rm u}
 -\frac{ T^{3/2}}{\sqrt{2\pi}c}F_{1/2}^{\rm p}
 +\frac{T^{5/2}}{2\sqrt{ \pi}c^3}F_{3/2}^{\rm u}
 +\frac{T^{5/2}}{4\sqrt{2\pi}c^3}F_{3/2}^{\rm p}.
\end{eqnarray}
\begin{figure}[t]
 \begin{center}
 \includegraphics[width=0.9\linewidth]{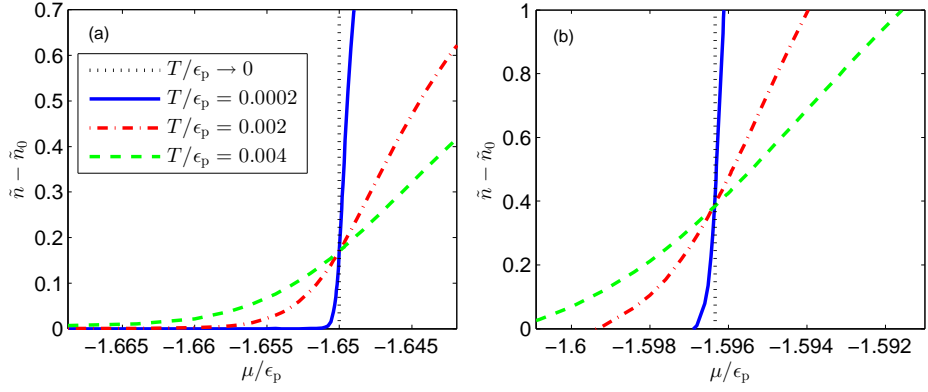}
 \caption{\label{crity0h55-n} (Color online)
 Universal scaling behaviour of the density $n$ at the quantum criticality for $h/c^2=0.55$:
 (a) at the  phase transition  V-FF and (b) at the phase transition  FF-FFLO.}
 \end{center}
\end{figure}

This spin-3/2 Fermi gas exhibiting rich quantum phase transitions
provides an ideal model to investigate quantum criticality of large
spin interacting fermions.
The phase diagram Fig.~\ref{fig:phase} has multifold critical
points.
In particular, the chemical potential and external field could drive
the system from one phase to another as these parameter pass across
the phase boundaries in the phase diagram Fig.~\ref{fig:phase} at
$T=0$.
All  phase transitions in  Fig.~\ref{fig:phase} are of the second
order.
Near the critical point, thermodynamical properties evolve into
certain universal  scaling forms as temperature tends to zero.
Below a crossover temperature (see the upper panel of
Fig.~\ref{fig:QC} below) the quantum criticality of the Luttinger
liquid gives the   dynamical exponent $z=1$.
However, beyond the crossover  temperature, a non-relativistic
quadratic dispersion leads to the free fermion criticality with
$z=2$, where new band excitations are involved
\cite{Guan:2009,Schlottmann:2012,Guan-Ho,Sachdev}.

For  $\gamma\gg1$ and   $h\gg T\gg |\mu-\mu_{\rm c}|$, the density
and compressibility  satisfy  the universal scaling function
\begin{eqnarray}
 && n-n_{0,i}=T^{d/z+1-1/vz}{\cal F}_i[(\mu -\mu_{{\rm c}i})/T^{1/vz}],\nonumber
 \\[6pt]
 && \kappa^*-\kappa^*_{0,i}=T^{d/z+1-2/vz} {\cal G}_i[(\mu-\mu_{{\rm c}i})/T^{1/vz}],
\end{eqnarray}
where we find $d/z+1-1/v z=1/2$, $1/vz=1$ with $d=1$, critical
exponent $z=2$ and correlation length exponent $v=1/2$.
In the above equation, $i=1,2,3,4$ denote the four phase transition
boundaries of V-FF, V-FP, FF-FFLO and FP-FFLO.
For $i=1,2$, both the background density $n_{0,i}$ and the
compressibility $\kappa_{0,i}$ are zero. For $i=3,4$, we have
\begin{eqnarray}
&&  n_{0,3}=\frac{1}{2\pi} \alpha, \quad n_{0,4}=\frac{2}{\pi} \beta
 \Big(1-\frac1\pi\beta
 +\frac{1}{\pi^2}\beta^2\Big),\nonumber\\
&&  \kappa_{0,3}=\frac1{2\pi}\frac{1}{\alpha},\quad
 \kappa_{0,4}=\frac{2}{\pi}\frac{1}{\beta}\Big(1-\frac3\pi\beta +\frac{6}{\pi^2} \beta^2\Big), \nonumber\\
&& \alpha=\Big\{2\alpha_1(h-h_{\rm
P})\Big[1+\frac2{3\pi}(2\alpha_1(h-h_{\rm P}))^{\frac12}
\Big]\Big\}^{\frac12},
 \nonumber\\
&&  \beta=\Big\{2\alpha_1(h_{\rm
P}-h)\Big[1+\frac2{\pi}(2\alpha_1(h_{\rm P}-h))^{\frac12}
\Big]\Big\}^{\frac12},
\end{eqnarray}
where $h_{\rm P}=\alpha_2 \epsilon_{\rm p}/2$ is the magnetic field
at the four phases (V, FF, FP and FFLO) coexistence point.
In the above equations the scaling functions ${\cal F}_i(x)$ and
${\cal G}_i(x)$ are
\begin{eqnarray}
&& {\cal F}_i(x)=\Phi_i\frac{F_{-1/2}(x)}{2\sqrt{\pi }}, \quad {\cal
G}_i(x)=\Psi_i\frac{F_{-3/2}(x)}{2\sqrt{\pi }},\quad i=1,4,
\nonumber \\ && {\cal
F}_i(x)=\Phi_i\frac{F_{-1/2}(2x)}{\sqrt{2\pi}},\quad
 {\cal G}_i(x)=\Psi_i\frac{F_{-3/2}(2x)}{\sqrt{2\pi}}, \quad
 i=2,3,
\end{eqnarray}
where the coefficients
\begin{eqnarray}
&&\Phi_1=\Phi_2=\Psi_1=\Psi_2=1,\nonumber \\
 && \Phi_3=1-\frac{2}{\pi}\alpha
 +\frac{1}{\pi^2}\alpha^2
 ,\quad  \Phi_4=1-\frac{8}{\pi}\beta+\frac{17}{\pi^2}\beta^2,
 \nonumber\\
 && \Psi_3=1+\frac{1}{2\pi}\alpha-\frac{1}{2\pi^2}\alpha^2,
 \quad \Psi_4=1+\frac{2}{\pi}\beta-\frac{10}{\pi^2}\beta^2.
\end{eqnarray}

\begin{figure}[t]
\begin{center}
 \includegraphics[width=0.9\linewidth]{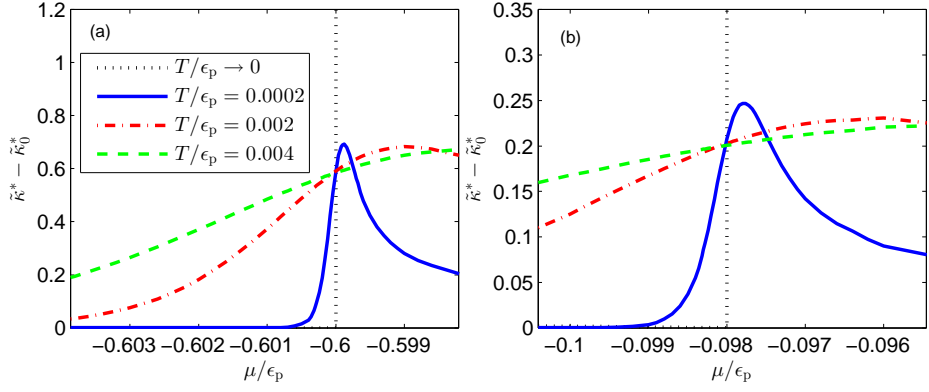}
 \caption{\label{crity0h05-kappa}
(Color online) Universal scaling behaviour of compressibility
$\kappa$ at the quantum criticality for  $h=0.05$: (a) at the  phase
transition  V-FP and (b) at the phase transition  FP-FFLO.}
\end{center}
\end{figure}

Let us define the dimensionless density $\tilde n=n_0 T^{-1/2}$ and
compressibility $\tilde \kappa^*=\kappa^* T^{1/2}$, then we have the
scaling forms
\begin{eqnarray}
\Delta\tilde n=\tilde n-\tilde n_{0,i}
 ={\cal F}_i(\Delta\mu/T), \quad
 \Delta\tilde \kappa^*=\tilde \kappa^*-\tilde\kappa_{0,i}
 ={\cal G}_i(\Delta\mu/T),
\end{eqnarray}
where $\Delta \mu=\mu-\mu_{{\rm c}i}$.
We observe that these physical quantities intersect at the  critical
point.
This intersection nature can be used to map out quantum criticality
of the model through the trapped gas at finite temperatures
\cite{Guan-Ho}, also see experimental study of the quantum
criticality for the  2D Bose atomic gases
\cite{Gemelke:2009,Hung:2010,Hung:2011a,Zhang:2012}.

We plot the critical properties of the density for the phase
transitions  V-FF and FF-FFLO  in  Fig. \ref{crity0h55-n}.
Whereas the critical properties of compressibility for the phase
transitions V-FP and FP-FFLO  are presented in Fig.
\ref{crity0h05-kappa}.
Universal scaling behaviours of the density and compressibility read
off the dynamical critical exponent $z=2$ and correlation critical
exponent $\nu=1/2$.

\begin{figure}[t]
\begin{center}
\includegraphics[width=0.7\linewidth]{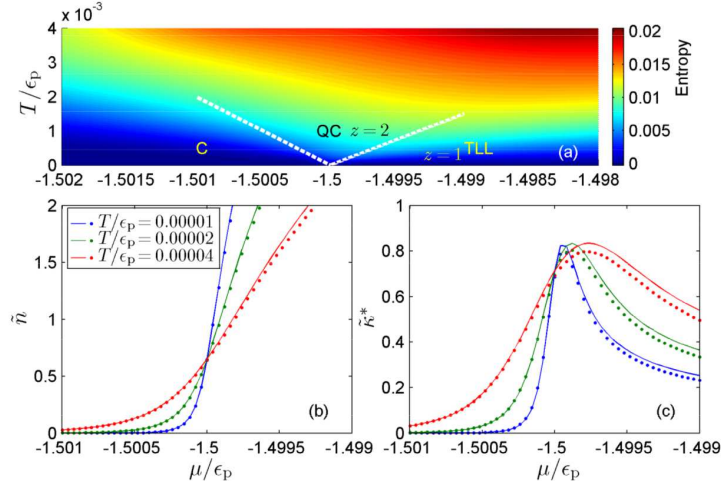}
\caption{\label{fig:QC} (Color online)  Upper panel: quantum
criticality of the phase transition  V-FFLO   near the critical
point $\mu_c=-1.5\epsilon_{\rm p}$ for the  fixed $h=\epsilon_{\rm
p}$ and $c>0$, where $C$ denote the classical region, QC means the
quantum critical regime and TLL is the Luttinger liquid. The white
dashed lines indicate the crossover temperature. Lower left (right)
panel: density (compressibility)  vs $\mu$ at different
temperatures. The solid lines denote  the analytical result from
(\ref{QC}) and the dot lines show  the numerical solution from the
TBA equations  (\ref{TBAE0}).  }
\end{center}
\end{figure}

In  Fig.~\ref{fig:QC} we further demonstrate the quantum criticality
near the phase transition from vacuum into FFLO  at the fixed
magnetic field $h=\epsilon_b$.
This phase transition involves a sudden change of the densities of
states of the spin-2  pairs and  single spin-3/2 atoms.
At  low temperatures, we obtain  the scaling functions
\begin{eqnarray}
&& \tilde{n} \approx \frac{1}{2\sqrt{\pi}}
 \left[F_{-\frac12}\Big(\frac{\mu-\mu_c}{T}\Big)
  +2^{\frac32}    F_{-\frac12}\Big(\frac{2(\mu-\mu_c)}{T}\Big)\right], \nonumber \\
&& \tilde \kappa^*\approx \frac{1}{2\sqrt{\pi}}
 \left[F_{-\frac32}\Big(\frac{\mu-\mu_c}{T}\Big)
  +2^{\frac52}    F_{-\frac32}\Big(\frac{2(\mu-\mu_c)}{T}\Big)\right], \label{QC}
\end{eqnarray}
where 
$\mu_c=-1.5\epsilon_{\rm p}$. Indeed we find that the scaling
functions of density and compressibility (\ref{QC}) read off the
dynamic critical exponent $z=2$ and correlation length exponent
$\nu=1/2$ by comparing with the universal forms given in
\cite{Fisher}.
In  Fig.~\ref{fig:QC}, we confirm these universal scaling forms
though numerical solution of the TBA equations (\ref{TBAE0}).
Similarly, entropy, magnetization, specific heat and susceptibility
can also map out the quantum criticality of the model.
%

\section{Conclusion}
\label{sc:c}

In conclusion, we have proposed a 1D integrable spin-3/2 fermionic gas of cold atoms with spin $SO(4)$ symmetry.
The symmetry, conserved quantities and integrability of this model has been
studied by means of the BA
We have shown that the integrable $SO(4)$ symmetry 
spin-3/2 Fermi gas exhibits  the  spin singlet and quintet
Cooper pairs in the two sets of $SU(2)\otimes SU(2)$ spin subspaces.
The Bethe ansatz equations and finite temperature thermodynamical equations
have been derived for the model in an analytical way.
In particular, using the Bethe ansatz exact solutions we have thoroughly
investigated spin pairing, full phase diagrams, equation of states, elementary
excitations, correlation functions, magnetism and quantum criticality of the
model.
We have also  shown that the $SO(4)$ symmetry Fermi gase possesses   various phases of
Luttinger liquids with novel a magnetism.
It is particular interesting that in the pure paired phase, breaking  a
2-string of $\nu$ leads to four spin-1 spinons in spin excitations.
The Luttinger liquid physics and and universal scaling behaviours of
thermodynamical properties in regard of  different spin states  provide insights
into understanding the  large spin phenomena in the 1D  interacting fermions.

Moreover, long distance asymptotics of various correlation functions have been
calculated by using conformal field theory.
In particular,  the FFLO-like pair-pair correlations exist in
the mixed phase of spin-2 pairs and single spin-3/2  atoms.
Furthermore, the density profiles of the trapped gas has been also discussed in
the context of a experimental setting with ultracold atoms.
It turns out that the trapping center can be a mixture of different types of
spin pairs  and unpaired single atoms accompanied by the multiple shells
involving the fully-polarized spin-2 pairs, the FFLO state, and the FF phase.
Our results open to further study of the non-$SU(\kappa)$ symmetry  interacting fermions of ultracold atoms in theory and experiment. 
\section*{Acknowledgement}

This work is supported by NSFC (Grant Nos. 91230203, 11174335,11304357,
11434013 and 11374331), CAEP, the National Basic Research
Program of China under Grant 2011CB922200 and 2012CB922101, and the
grant from Chinese Academy of Sciences. XWG acknowledges the Beijing
Computational Science Research Center and KITPC, Beijing  for their
kind hospitality. He  has been partially supported by the Australian
Research Council.

\end{document}